\documentclass[useAMS,usenatbib]{mn2e}
\usepackage{natbib}
\usepackage{color,graphicx} 
\usepackage{amsmath}        
\usepackage{amsfonts}       
\usepackage{amssymb}        
\usepackage{wasysym}        

\begin{document}
\author[K.~Liu et al.]
{K.~Liu,$^{1,2}$ R.~Karuppusamy,$^{2}$ K.~J.~Lee,$^{2}$
B.~W.~Stappers,$^{3}$ M.~Kramer,$^{2,3}$
\newauthor R.~Smits$^{4}$ M.~B.~Purver,$^{3}$ G.~H.~Janssen,$^{4,3}$ and D.~Perrodin,$^{5}$\\
  $^{1}$Station de radioastronomie de Nan\c{c}ay, Observatoire de
  Paris, CNRS/INSU, F-18330 Nan\c{c}ay, and Laboratoire de Physique et \\
  Chimie de l'Environnement et de l'Espace LPC2E CNRS-Universit\'{e} d'Orl\'{e}ans, F-45071 Orl\'{e}ans Cedex 02, France \\
  $^{2}$Max-Planck-Institut f\"{u}r Radioastronomie, Auf dem H\"{u}gel
  69, D-53121 Bonn, Germany \\
  $^{3}$University of Manchester, Jodrell Bank Centre of Astrophysics,
  Alan-Turing Building, Manchester M13 9PL, UK\\
  $^{4}$ ASTRON, Oudehoogeveensedijk 4, Dwingeloo, 7991PD, The Netherlands\\
  $^{5}$INAF - Osservatorio Astronomico di Cagliari, Via della Scienza 5, 09047 Selargius (CA),
  Italy\\
  }

\title[Profile instability of PSR~J1022+1001]{Single pulse and profile variability study of PSR~J1022+1001}

\maketitle

\begin{abstract}
Millisecond pulsars (MSPs) are known as highly stable celestial
clocks. Nevertheless, recent studies have revealed the unstable
nature of their integrated pulse profiles, which may limit the
achievable pulsar timing precision. In this paper, we present a case
study on the pulse profile variability of PSR~J1022+1001. We have
detected approximately 14,000 sub-pulses (components of single
pulses) in 35-hr long observations, mostly located at the trailing
component of the integrated profile. Their flux densities and
fractional polarisation suggest that they represent the bright end
of the energy distribution in ordinary emission mode and are not
giant pulses. The occurrence of sub-pulses from the leading and
trailing components of the integrated profile is shown to be
correlated. For sub-pulses from the latter, a preferred pulse width
of approximately 0.25\,ms has been found. Using simultaneous
observations from the Effelsberg 100-m telescope and the Westerbork
Synthesis Radio Telescope, we have found that the integrated profile
varies on a timescale of a few tens of minutes. We show that
improper polarisation calibration and diffractive scintillation
cannot be the sole reason for the observed instability. In addition,
we demonstrate that timing residuals generated from averages of the
detected sub-pulses are dominated by phase jitter, and place an
upper limit of $\sim700$\,ns for jitter noise based on continuous
1-min integrations.
\end{abstract}


\section{Introduction} \label{sec:intro}
Rotation-powered pulsars have been shown to be stable celestial
clocks \citep[e.g.][]{vbc+09} and thus excellent tools in the
ongoing endeavours of high precision gravity tests \citep{ksm+06},
neutron star equations of state \citep{dpr+10,opr+10}, and
gravitational wave detection \citep{vlj11,ych+11,dfg13,src+13}.
These experiments utilise the high precision timing of
millisecond pulsars (MSPs), which are known for their rotational
stability and the stability of their integrated pulse profiles
(which do not change over many years). However, recent studies have
shown that the profiles of MSPs can exhibit low-level instabilities
that may limit the achievable timing precision. In general, these
instabilities can be categorized into two types: temporally
uncorrelated variabilities caused by the instability of single
pulses \citep[e.g.][]{jak+98,lkl+11,sc12}, and temporally correlated
variations induced by other phenomena \citep[e.g.][]{kxc+99,es03a}.
The contribution of jitter to the pulsar timing noise could be
mitigated by simply increasing the exposure time or by correcting
the biased integrated profiles with suitable algorithms
\citep{cs10,ovd+13}. Comprehensive studies of profile instabilities
would give new insights into the fundamental timing limits and
thereby potentially improving the timing precision, which is
essential for the aforementioned astrophysical experiments.

PSR~J1022+1001 is a MSP with a rotation period of approximately
16\,ms. It exhibits a regular rotational behaviour and is included
in the current pulsar timing array observations for the purpose of
gravitational wave detection \citep[e.g.][]{man13}. The pulsar is in
a binary with a 7.8-day orbital period with a white dwarf companion,
and is thus a potential laboratory for testing the Strong
Equivalence Principle \citep[e.g.][]{fwe+12,afw+13}. The integrated
pulsed profile of the pulsar consists of a double-peaked structure
at L-band frequencies with a highly linearly polarised trailing
component. The amplitude ratio of the two components has been shown
to change significantly as a function of frequency \citep{rk03}.
There is also evidence that the ratio is unstable across time and
may even evolve on short timescales of tens of minutes
\citep{kxc+99,pur10}. Meanwhile, as the trailing component of the
pulse profile is highly polarised, profile instability could also
arise from improper polarisation calibration due to an imperfect
receiver model \citep{hbo04}. A better understanding of the profile
variation in this system is therefore essential in order to further
improve the timing precision of PSR~J1022+1001. This may be
used to assess and perhaps correct for profile variations in other
MSPs.

Examining single pulses has not been common in MSPs owing to their
typically lower flux densities compared to normal pulsars.
Exceptions are the so-called giant pulses which have been
detected only in a few bright MSPs
\citep[e.g.][]{cst+96,rj01,jkl+04,kbmo05}. Analysis of the weak
single pulses has been only carried out in the few brightest
MSPs \citep{jak+98,jap01,es03a,sc12,ovb+14}. Based on an
autocorrelation analysis, \cite{es03a} have shown evidence for pulse
intensity modulation at the phase of the trailing component in
PSR~J1022+1001. However, no detailed investigation into the single
pulses has been undertaken so far, due to either limited system
sensitivity or the lack of high-resolution instrumentation.

The rest of this paper is structured as follows: In
Section~\ref{sec:obs} we describe the details of observations and
data processing. Results of the detected single pulses, integrated
profile stability, and timing analysis are presented in
Section~\ref{sec:res}. We conclude and discuss the results in
Section~\ref{sec:conclu}.

\section{Observation} \label{sec:obs}
Simultaneous observations of 7 to 9 hours in duration of
PSR~J1022+1001 were conducted with the Westerbork Synthesis
Radio Telescope (WSRT) and the Effelsberg 100-m Radio Telescope, at four epochs. A summary of the observations can be found in
Table~\ref{tab:obs}. The two telescopes were chosen as they provide
similar sensitivities at L-band so that profile instability could be
verified from both sites at the same time. In addition, Effelsberg
is an alt-azimuth telescope, meaning that it tracks a source
across the sky, the changing parallactic angle (PA) results in
varied Stokes parameters which need to be properly calibrated. On the other hand, the WSRT is equatorially mounted with no PA
variation and thus in principle not influenced by this effect.
Therefore, using data from these two telescopes allows us to
cross-check the result of polarisation calibration which was
claimed to be the main cause of profile instability
\citep{hbo04,van13}.

\begin{table}
\centering \caption{Parameters of the simultaneous observations of
PSR~J1022+1001 with the WSRT and the Effelsberg 100-m Radio
Telescope. The symbols $T$, $f_{\rm o}$, and $\Delta t_{\rm s}$
represent the duration of observation, overlapping observing
frequency range, and the single pulse time resolution at the
WSRT.} \label{tab:obs}
\begin{tabular}[c]{cccc}
\hline
MJD           & $T$ (hr) &$f_{\rm o}$ (MHz) &$\Delta t_{\rm s}$ ($\mu$s) \\
\hline
55909         & 7.0      &1300-1440         &4.0\\
55974$^\star$ & 8.4      &1300-1440         &4.0\\
56158 & 8.5   &1300-1440         &2.0\\
56257 & 9.0   &1300-1365$^*$     &2.0\\
\hline
\end{tabular}
\\$^\star$At this epoch the Effelsberg data were corrupted due to a problem with the signal
attenuators (see Appendix~\ref{sec:app} for more details).
\\$^*$At this epoch three 25\,MHz sub-bands of Effelsberg data
were lost due to hard drive failure.
\end{table}

At the WSRT we used the PuMa-II system to record 8-bit baseband data
\citep{ksv08}, which were later processed offline in two
stages. In the first stages the 8$\times$20\,MHz bands were combined
to form a multi-band data stream corresponding to a 160\,MHz
contiguous band centered at 1380\,MHz, at the original time
resolution of 25\,ns. The DSPSR software \citep[for details
see][]{vb11} was used to coherently dedisperse the data over the
entire bandwidth using a 64-channel synthetic software filterbank at
the DM of 10.246\,cm$^{-3}$pc and folded to form 10-s averages in
the second stage. Additionally, we searched for significant
single pulses at this stage. This was done after summing up the
total intensities of all frequencies, and averaging by the number of
frequency channels. Any pulse period that contained a peak greater
than five times the root-mean-square (rms) of the off-pulse noise
was written out as a single pulse candidate. In the post-processing
stage, these were then confirmed to be true pulses by comparing the
phase to that of the integrated pulse profile.

At Effelsberg, the new PSRIX pulsar backend (Karuppusamy et al., in
prep) was used in baseband mode. These data were reduced in the
same way as the WSRT data, except that here, 8$\times$25\,MHz bands
were produced with a central frequency of 1347.5\,MHz. In addition,
at regular intervals of 40\,min, the telescope was pointed
$0.5^{\circ}$ off-source for 90\,s, when a pulsed noise signal
was injected at $45^{\circ}$ into the feed probes as a calibrator
for polarimetry.

The 10-s integrations and single pulses were then processed with the
\textsc{psrchive} software package \citep{hvm04}. For the 10-s
integrations from Effelsberg we performed polarisation calibration
with the single-axis receiver model which corrects for the
differential gains between the two
feeds\footnote{\textbf{http://psrchive.sourceforge.net/manuals/pac/}}.
Next we used \textsc{pazi (psrchive's interactive RFI zapper)} to
clean the radio frequency interference (RFI) by visual examination
of the averaged pulse profiles and single pulse candidates. Finally,
we selected all 10-s integrations that overlapped in time between
the two telescopes, and then removed the non-overlapping frequency
channels, leading to an overlapping observing frequency range
of $1300-1440$\,MHz.

\begin{figure}
\centering
\includegraphics[scale=0.34,angle=-90]{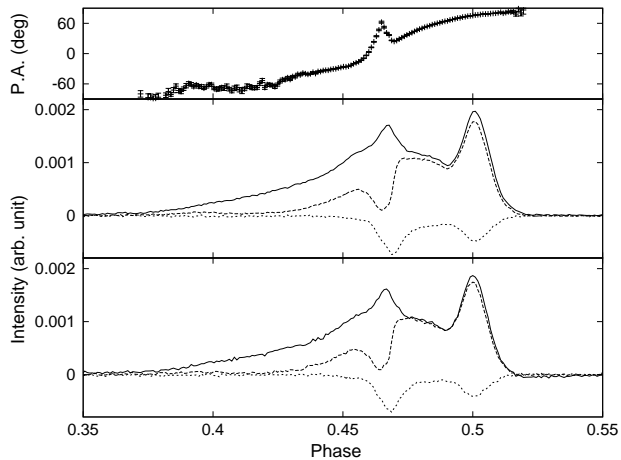}
\caption{Polarisation profiles of PSR~J1022+1001 obtained at
MJD~56257 from the Effelsberg (middle) and the WSRT (bottom)
observation at 1.3\,GHz. The P.A. in the top panel stands for the
polarisation position angle of the linear component. The Effelsberg
and the WSRT data show consistent P.A. curves and here we plot the
one from the Effelsberg data as a demonstration. The solid, dashed
and dotted lines represent the Stokes parameters $I$, $L$ and $V$,
respectively. There is a ``negative'' dip, i.e., an underestimation
of system temperature, on the WSRT total intensity profile (at phase
0.52) which corresponds to the low-bit sampling artefact discussed
in \citet{ja98}. This is due to the fact that the signal was
originally quantized using 2 bits per sample at each individual dish
of the WSRT. Further investigation shows that the dip is mostly from
high frequency channels within each 20\,MHz band. The total
intensity profile is slightly lowered by this underestimation, which
makes the valley region between the two components appear over
polarised. Further investigations can be found in
Appendix~\ref{sec:app}. \label{fig:prof-cal}}
\end{figure}

Example polarisation profiles of PSR~J1022+1001 from both the WSRT
and the Effelsberg observation at MJD~56257 centred at 1320\,MHz,
can be found in Fig.~\ref{fig:prof-cal}. The integrated profile
shows a double-peaked structure, with the trailing component
consisting of highly linearly polarised emission. The polarisation
properties are in agreement between the two sites, with a
correlation coefficient of 0.999 \citep[as defined in][]{lkl+11}.
Given the profile instability of this pulsar, especially in
observations separated by several years \citep{pur10}, the
polarisation components qualitatively agree with the previous
observations \citep{kxc+99,rk03,van13}.

\section{Results} \label{sec:res}
\subsection{Sub-pulse properties} \label{ssec:sgl}
\begin{figure}
\centering
\includegraphics[scale=0.35,angle=-90]{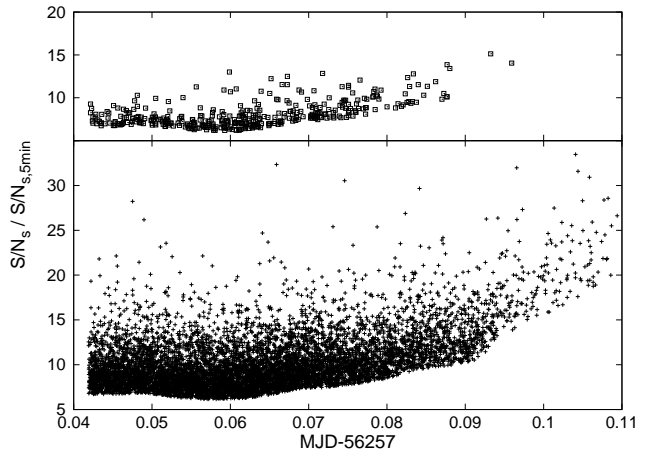}
\caption{Sub-pulses detected at MJD~56257 as a function of time and
their peak signal-to-noise ratios (S/N$_{\rm s}$) divided by
the local average S/N of all single pulses observed during the 5-min
interval centred on the epoch of the single pulse ($\langle{\rm
S/N}_{\rm s}\rangle_{\rm 5min}$). The location of peak bin was
selected based on a 16-$\mu$s time resolution. The top panel shows
detections at the phase of the leading component in the averaged
profile, and the bottom shows those at the phase of the trailing
one. \label{fig:sgl-26Nov2012}}
\end{figure}

In a total of 35 hours of observations, we have approximately
14,400 single pulse detections above our 5-$\sigma$ threshold, which
corresponds to about 700 sub-pulses\footnote{In each rotation, the
pulsar produces a single pulse which can be composed of one or more
sub-pulses. This better reflects our situation as occasionally, more
than one sub-pulses can be detected from a single period.} located
at the leading component of the integrated profile and about 13,700
at the trailing component\footnote{The leading and trailing
component pulses are distinguished by an arbitrary bound of
rotational phase 0.485 in Fig.~\ref{fig:prof-cal}.}. The detections
were all made from the observations at MJD~55909 and 56257 as the
pulsar was significantly weaker due to interstellar scintillation at
MJD~55974 and 56158. The highest peak signal-to-noise ratio of
detection is close to 10. Fig.~\ref{fig:sgl-26Nov2012} shows the
sub-pulse detections achieved at MJD~56257 at their occurrence
epochs and peak signal-to-noise ratios (S/N$_{\rm s}$) relative to
the local average S/N of all single pulses observed during the
5-min interval centred at the epoch of the single pulse
($\langle{\rm S/N}_{\rm s}\rangle_{\rm 5min}$). We can see that
there is an evolution in the ratio which is due to the pulsar flux
decreasing because of interstellar scintillation. With our
sensitivity, useful detections are feasible only when integrating
5-min observations results in a profile of peak S/N (S/N$_{\rm
5min}$) above 30. The detection rate is highest between
MJD~56257.055 and 56257.060, corresponding to 4.5\% of all
rotations. We have calculated time separations between neighbouring
sub-pulses at the trailing component within this period of time, and
over 95\% of them are less than 1\,s, indicating that pulse nulling
with duration longer than 1\,s ($\simeq60$ rotations) is not
detected.

\begin{figure}
\centering
\includegraphics[scale=0.35,angle=-90]{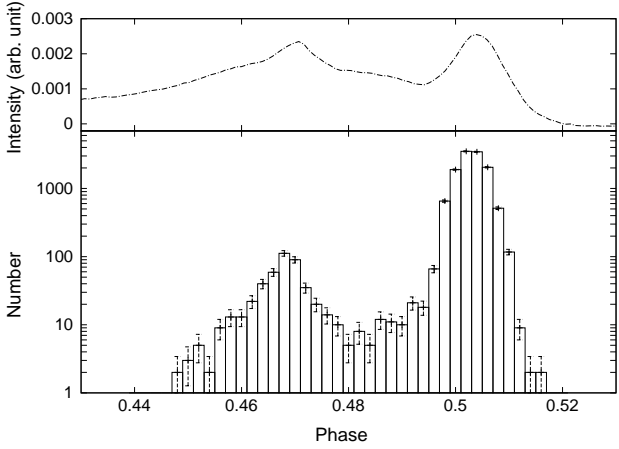}
\caption{Distribution of peak amplitude phase for all detected
sub-pulses (lower panel), in comparison with the integrated profile
(upper panel). Again, the peak bins were chosen based on a 16-$\mu$s
time resolution. The integrated profile is from the WSRT observation
at MJD~56257. \label{fig:pha}}
\end{figure}

\begin{figure}
\centering
\includegraphics[scale=0.5,angle=-90]{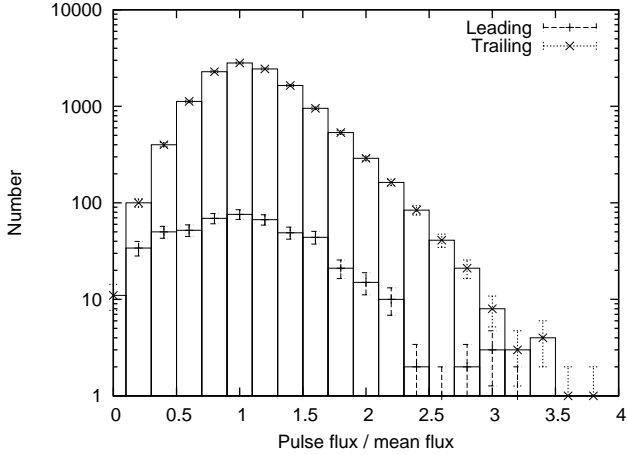}
\caption{Distributions of relative flux densities to averaged single
pulse flux densities in 5-min integrations, for sub-pulses detected
at the leading and trailing components, respectively. No pulse
with flux above 3.8 times the average has been seen.
\label{fig:powspec}}
\end{figure}

Fig.~\ref{fig:pha} shows the distribution of the peak amplitude
phases for the detected sub-pulses, compared with the
integrated profile. All detections were obtained within the on-pulse
phase and most of them are clustered within the range defined by the
trailing component in the average profile. The two maxima correspond
to the two peaks in the phase-resolved modulation index of
\cite{es03}. In Fig.~\ref{fig:powspec} we plot the distribution of
the flux densities of the sub-pulses divided by the averaged flux
densities of a single pulse in 5-min integrations. The plot
demonstrates that none of the detections has a flux density
that is above 3.8 times the average, and there is
therefore no evidence for giant pulses.

\begin{figure}
\centering
\includegraphics[scale=0.35,angle=-90]{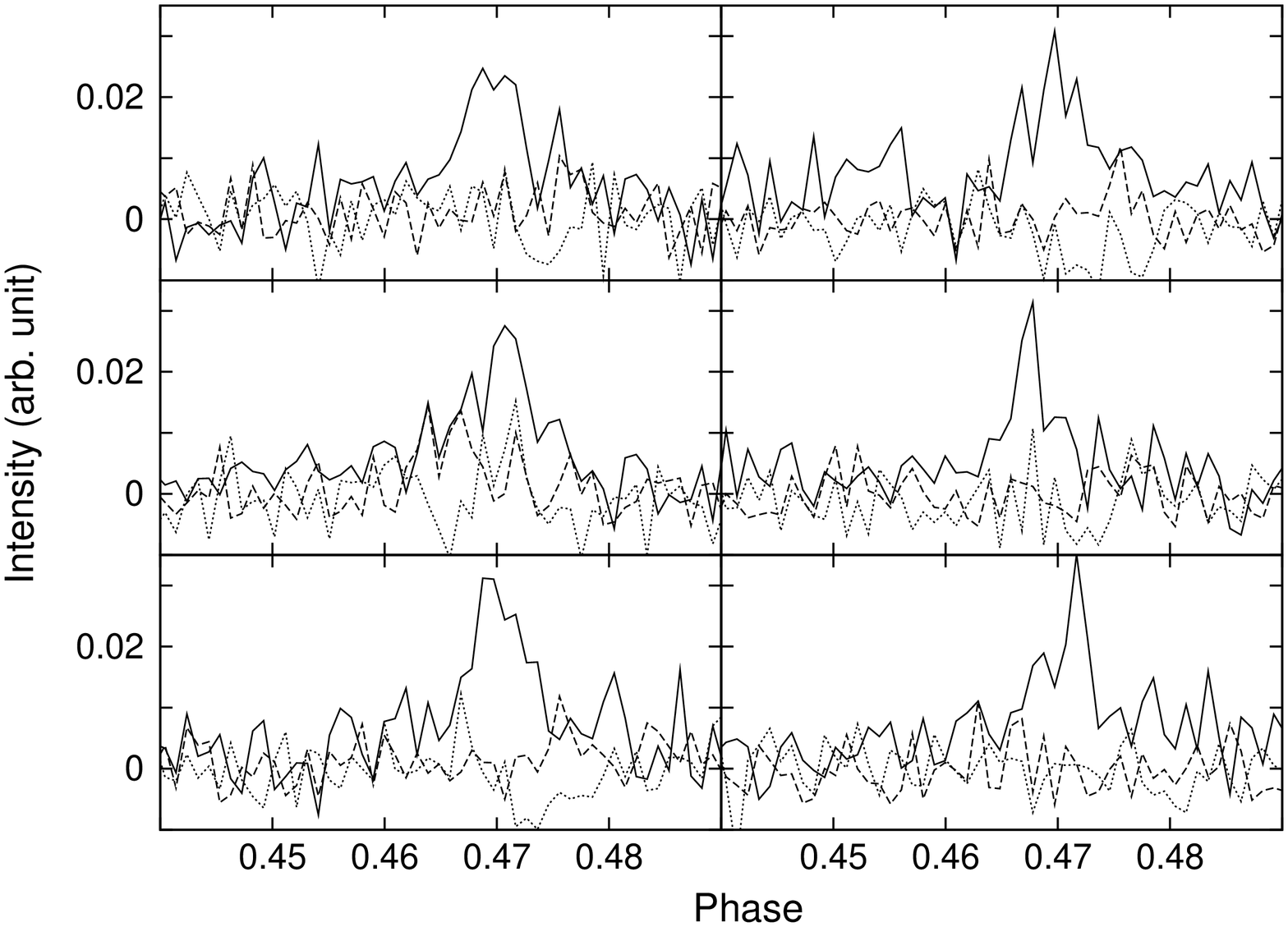}
\includegraphics[scale=0.35,angle=-90]{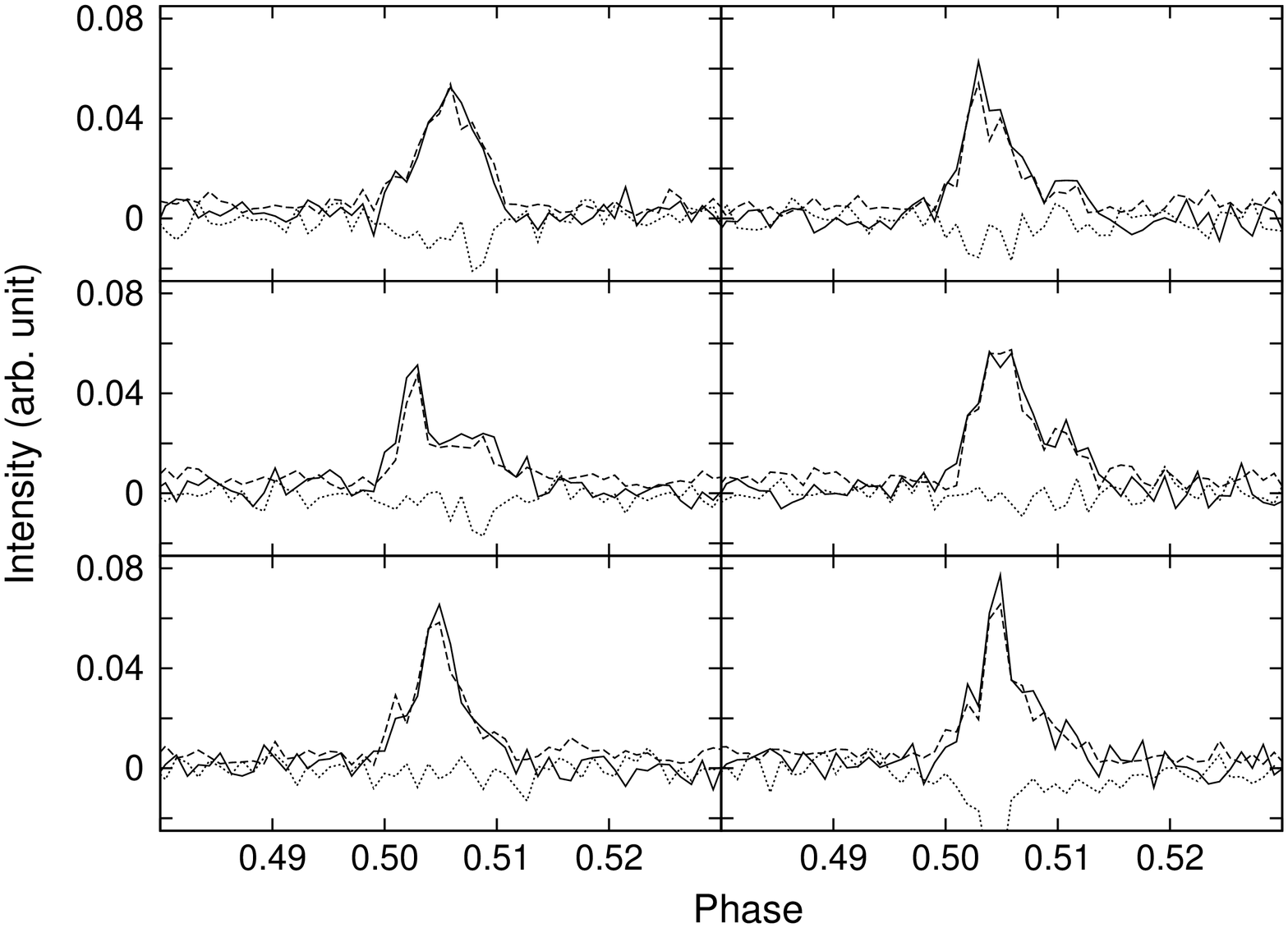}
\includegraphics[scale=0.35,angle=-90]{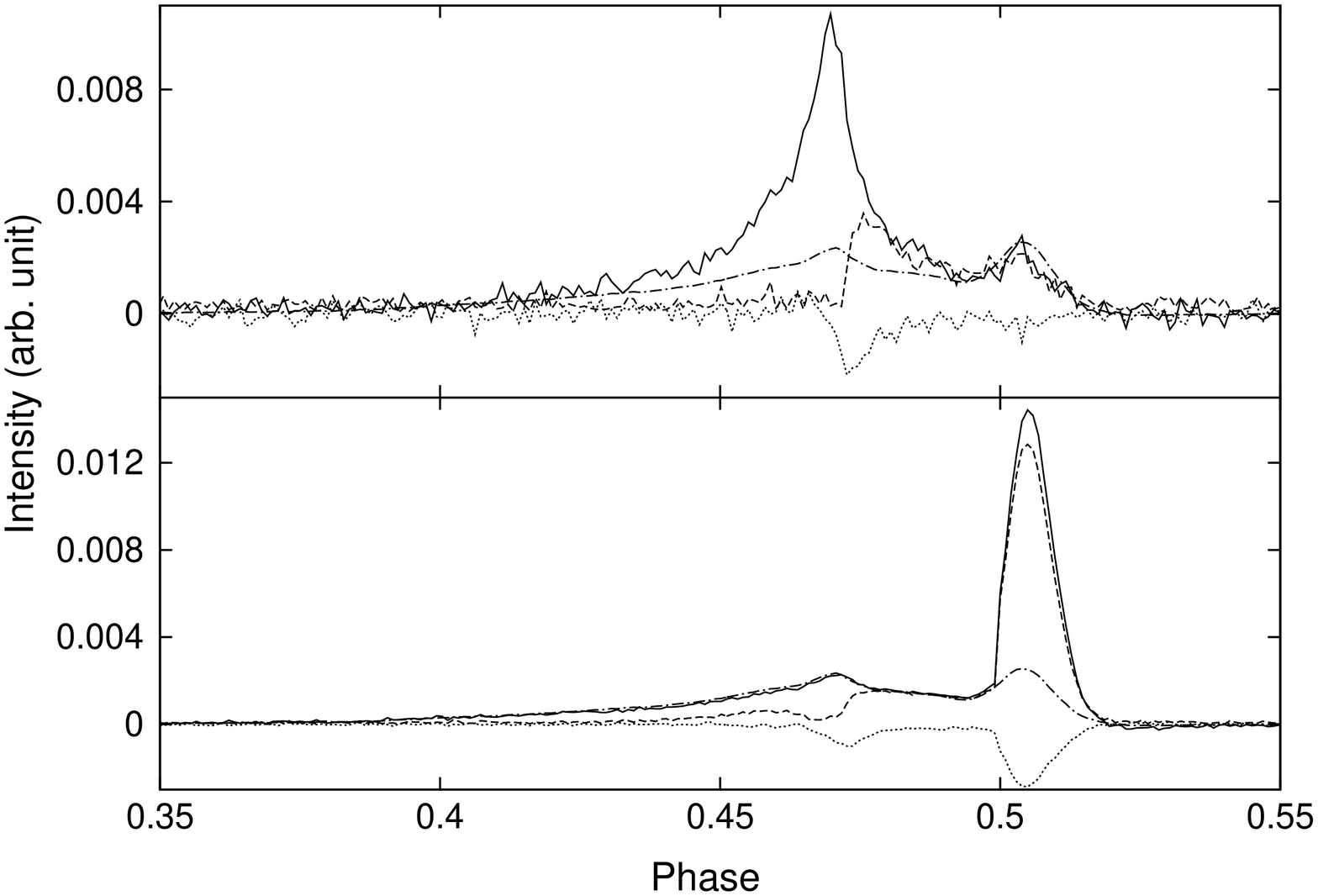}
\caption{Example polarisation profiles of sub-pulses detected at
MJD~56257, at the phase of the leading (top) and trailing (middle)
component in the average profile. The bottom plot shows the
integrations of detected sub-pulses from the leading (upper panel)
and trailing (lower panel) components, respectively. The solid,
dashed and dotted lines stand for the Stokes $I$, $L$ and $V$,
respectively. The dashed-dotted line represents the averaged total
intensity of all periods. \label{fig:sample}}
\end{figure}

Fig.~\ref{fig:sample} presents example polarisation profiles of
sub-pulses at the phase of the leading and trailing components. It can be seen that pulses from the leading component
do not show significant polarisation, while those from the trailing
component are highly linearly polarised. The polarisation
fractions of both components are consistent with the integrated
pulse profile, suggesting that the detected sub-pulses
represent the bright end of the energy distribution of all pulses.
The bottom plot in Fig.~\ref{fig:sample} shows the averages of the
leading and trailing component pulses\footnote{Here we selected
pulses with 5$-\sigma$ detection in only one component.
Occasionally, bright emission can occur at both components within a
single period, which will be discussed in the next paragraph.},
respectively, whose polarisation characteristics are in agreement
with the corresponding individual detections. It is interesting to
note that in both averages, the components that do not include
detected sub-pulses are consistent with the average of all periods
in total intensity. This means that during the occurrence of bright
emission in one component, there still exists emission at an average
level at the location of the other component.

\begin{figure}
\centering
\includegraphics[scale=0.47,angle=-90]{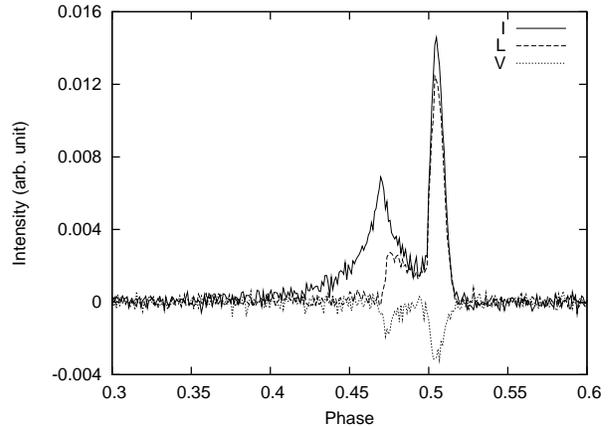}
\caption{Polarisation profile averaged over all detected bi-pulses,
where sub-pulses were detected at both components within a single
period. \label{fig:1-2}}
\end{figure}

Meanwhile, sub-pulses can be detected from the leading and trailing
components within the same pulse period. In total, about 280 such
events have been detected. In all events, the two phases of peak
amplitude are well separated from the phase bound 0.485 (used to
group the detections), excluding detections of a single pulse with
broad width spreading over the region of both components. We will
later refer to such an event as a `bi-pulse', and their
corresponding integrated profile is shown in Fig.~\ref{fig:1-2}. The
valley region between the two components is significantly steeper in
this profile than the ordinary average in Fig.~\ref{fig:prof-cal},
while the polarisation properties of the two components are
consistent. To investigate the correlation of occurrence between the
leading and trailing component pulses, we carried out a further
statistical study. Pulses were selected at both epochs, from the
period of time when the leading component pulses could be detected.
A summary of the selections can be found in Table~\ref{tab:1+2}. We
formed the statistical test problem in the following statement:
\begin{equation}
    \left\{ \begin{array}{l l}
        H_0 & \textrm{Leading and trailing pulses are uncorrelated.}\\
        H_1 & \textrm{Leading and trailing pulses are correlated.}
    \end{array}\right. \nonumber
\end{equation}
The bi-pulse phenomenon is defined as: both the leading and trailing
component pulses are detected within a single period. Thus, under
the null-hypothesis $H_0$, in each pulse period the probability of
observing bi-pulses is
\begin{equation}
    P_{\rm b}=P_{\rm l} P_{\rm t}\,,
\end{equation}
where $P_{\rm b}, P_{\rm l}$ and $P_{\rm t}$ are the
probabilities of observing bi, leading, and trailing component
pulses, respectively. By combining the statistics from these two
epochs, we have calculated the observed $P_{\rm l}$ and $P_{\rm t}$
which gives an expected value of $P_{\rm b}$ as
$3.80\pm0.16\times10^{-5}$ under the null-hypothesis $H_0$.
Meanwhile, the observed bi-pulse occurrence probability was
calculated to be $6.32\pm0.39\times10^{-4}$, about 17 times larger.
For a more quantitative study, over $N$ pulse periods the
statistical distribution function $f$ of the total number of
bi-pulses, leading and trailing component pulses are
\begin{eqnarray}
    f(N_{\rm b}|P_{\rm b},N)=C_{N}^{N_{\rm b}} P_{\rm b} ^{N_{\rm b}} (1-P_{\rm
    b})^{N-N_{\rm b}}\,,\\
    f(N_{\rm l}|P_{\rm l}, N)=C_{N}^{N_{\rm l}} P_{\rm l} ^{N_{\rm l}} (1-P_{\rm
    l})^{N-N_{\rm l}}\,,\\
    f(N_{\rm t}|P_{\rm t}, N)=C_{N}^{N_{\rm t}} P_{\rm t} ^{N_{\rm t}} (1-P_{\rm
    t})^{N-N_{\rm t}}\,,
\end{eqnarray}
where $C_{b}^{a}$ is the binomial coefficient defined as
$C_{b}^{a}=(b!)/(a!(b-a)!)$. Using the Bayesian theorem with a flat
prior for the distributions of probability of each type of pulses,
we have
\begin{eqnarray}
    f(P_{\rm l}|N_{\rm l}, N)=f(N_{\rm l}|P_{\rm l}, N)(N+1)\,,\\
    f(P_{\rm t}|N_{\rm t}, N)=f(N_{\rm t}|P_{\rm t}, N)(N+1)\,,
\end{eqnarray}
where $f(P_{\rm l}|N_{\rm l}, N)$ and $f(P_{\rm t}|N_{\rm t}, N)$
are the posterior for $P_{\rm l}$ and $P_{\rm t}$. Thus the
probability distribution of $N_{\rm b}$ given $N_{\rm l}$ and
$N_{\rm t}$ is
\begin{eqnarray}
    f(N_{\rm b}|N_{\rm l},N_{\rm t}, N)&=&\int_{0}^{1} \int_0^1 dP_{\rm l}
    dP_{\rm t} C_{N}^{N_{\rm b}} (P_{\rm l} P_{\rm t}) ^{N_{\rm b}} \nonumber
    \\
    &&(1-P_{\rm l} P_{\rm t})^{N-N_{\rm b}} \nonumber \\
    &&f(P_{\rm l}|N_{\rm l}, N) f(P_{\rm t}|N_{\rm t}, N)\,.
\end{eqnarray}
Under the null hypothesis $H_0$, we have
\begin{equation}
    P(N_{\rm b} >N_{\rm B} \textrm{ or } N_{\rm b}<N_{\rm A}
    )=1-\sum_{N_{\rm A}}^{N_{\rm B}}f(N_{\rm b}|N_{\rm l},N_{\rm t}, N)\,,
\end{equation}
where $P(N_{\rm b} >N_{\rm B}~\textrm{or}~N_{\rm b}<N_{\rm A})$ is
the probability of finding that the occurrence of bi-pulses
$N_{\rm b}$ is greater than the threshold $N_{\rm B}$ or less than
the threshold $N_{\rm A}$, under the assumption that the leading and
trailing component pulses are uncorrelated. This provides a good
statistic for testing whether the observations agree with the
null hypothesis $H_0$. Accordingly, we have estimated the
probability to detect more than $N_{\rm b}$ bi-pulses
($P(n>=N_{b})$) based on the statistics in Table~\ref{tab:1+2},
which appears to be very close to zero at both epochs. We therefore
conclude that the leading and trailing component pulses are
correlated in their occurrence, otherwise it would be extremely rare
to observe such a high value of $N_{\rm b}$ with the null-hypothesis
$H_0$. In young pulsars, correlated emission modulation at different
rotational phases has been found in a few cases \citep[][Karuppusamy
et al. in prep.]{wwj12}.

\begin{table}
\centering \caption{Summary of selected pulses used to investigate
the occurrence correlation between the leading and trailing
component pulses. Here N, N$_{\rm l}$, N$_{\rm t}$ and N$_{\rm b}$
denote the number of periods within the time range the pulses were
chosen from, number of leading and trailing component pulses, and
number of bi-pulses, respectively. The probability of detecting more
than $N_{\rm b}$ bi-pulses is denoted by $P(n\geqslant N_{b})$}.
\label{tab:1+2}
\begin{tabular}[c]{ccc} \\
\hline
Phase         &MJD~55909       &MJD~56257\\
\hline
N            &$2.1\times10^5$  &$2.0\times10^5$ \\
N$_{\rm l}$   &72               &525            \\
N$_{\rm t}$   &2737             &7929           \\
N$_{\rm b}$   &25               &234            \\
$P(n>=N_{b})^{\ast}$ &$10^{-252}$ &$10^{-881}$\\
\hline
\end{tabular}
\\$^*$ Due to the difficulties in the numerical evaluation of the
integral, we used asymptotic values for the error function. This
could result in one order of magnitude error, but will not change
the conclusion.
\end{table}

\begin{figure}
\centering
\includegraphics[scale=0.55,angle=-90]{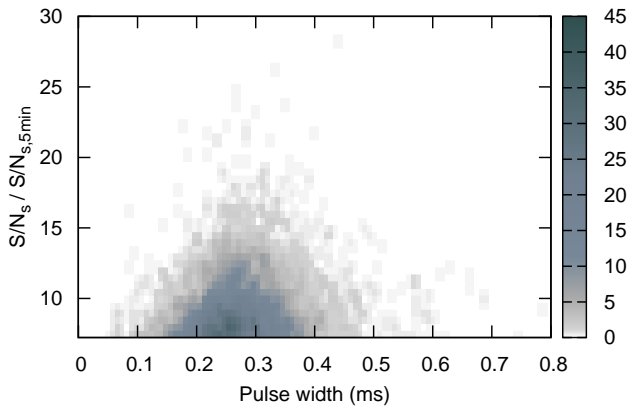}
\caption{Distribution of pulses from the trailing component by their
relative S/N$_{\rm s}$ and width. To avoid bias caused by the
time-variant detection threshold, only pulses with epoch between
MJD~56257.042 and 56257.066, and $({\rm S/N}_{\rm
p})/(\langle{\rm S/N}_{\rm s}\rangle_{\rm 5min})>7$, were
selected. Again, the peak bins were selected based on a 16-$\mu$s
time resolution. \label{fig:widvf}}
\end{figure}

\begin{figure}
\centering
\includegraphics[scale=0.55,angle=-90]{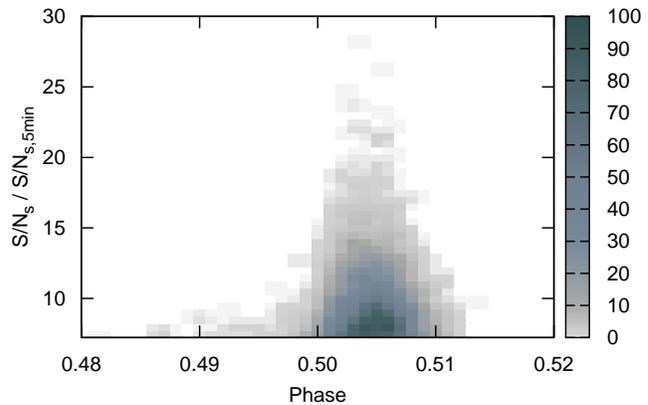}
\caption{Distribution of the same pulses as in Fig.~\ref{fig:widvf}
on a phase$-$relative S/N$_{\rm s}$ plane. \label{fig:phavf}}
\end{figure}

\cite{jak+98} showed an inverse correlation between pulse width and
peak amplitude for the sub-pulses of PSR~J0437$-$4715. In
Fig.~\ref{fig:widvf} we perform the same investigation by showing
the distribution of detected pulses from the trailing component on a
width$-$relative S/N$_{\rm s}$ plane. Here, to avoid bias caused by
a variable detection threshold, we only selected detections
between MJD~56257.042 and 56257.066 and with $({\rm S/N}_{\rm
s})/(\langle{\rm S/N}_{\rm s}\rangle_{\rm 5min})>7$. It is
interesting to note that, instead of an anti-correlation, there
exists a favoured pulse width\footnote{Here the pulse width is
defined by the on-pulse region.} value of about 0.25\,ms which is
consistent for pulses of different relative S/N$_{\rm s}$ values.
The probability of occurrence also does not favour pulses with
narrow width and high peak amplitude. Fig.~\ref{fig:phavf} gives the
distribution of the same pulses with respect to their phases and
relative S/N$_{\rm s}$. The plot demonstrates that the occurrence
phases of the sub-pulses are strongly clustered, especially for
those of peak 10 times above average. The phenomenon that brighter
pulses show up earlier in phase, found for PSR~J1713+0747 in
\cite{sc12}, is not seen. The property of regular shape and phase
could possibly be translated into precise pulse time-of-arrival
(TOA) measurements, which will be further investigated in
Section~\ref{ssec:timing}.

\subsection{Variation of integrated profiles and sub-pulse occurrence} \label{ssec:prof stab}
\begin{figure}
\centering
\includegraphics[scale=0.48,angle=-90]{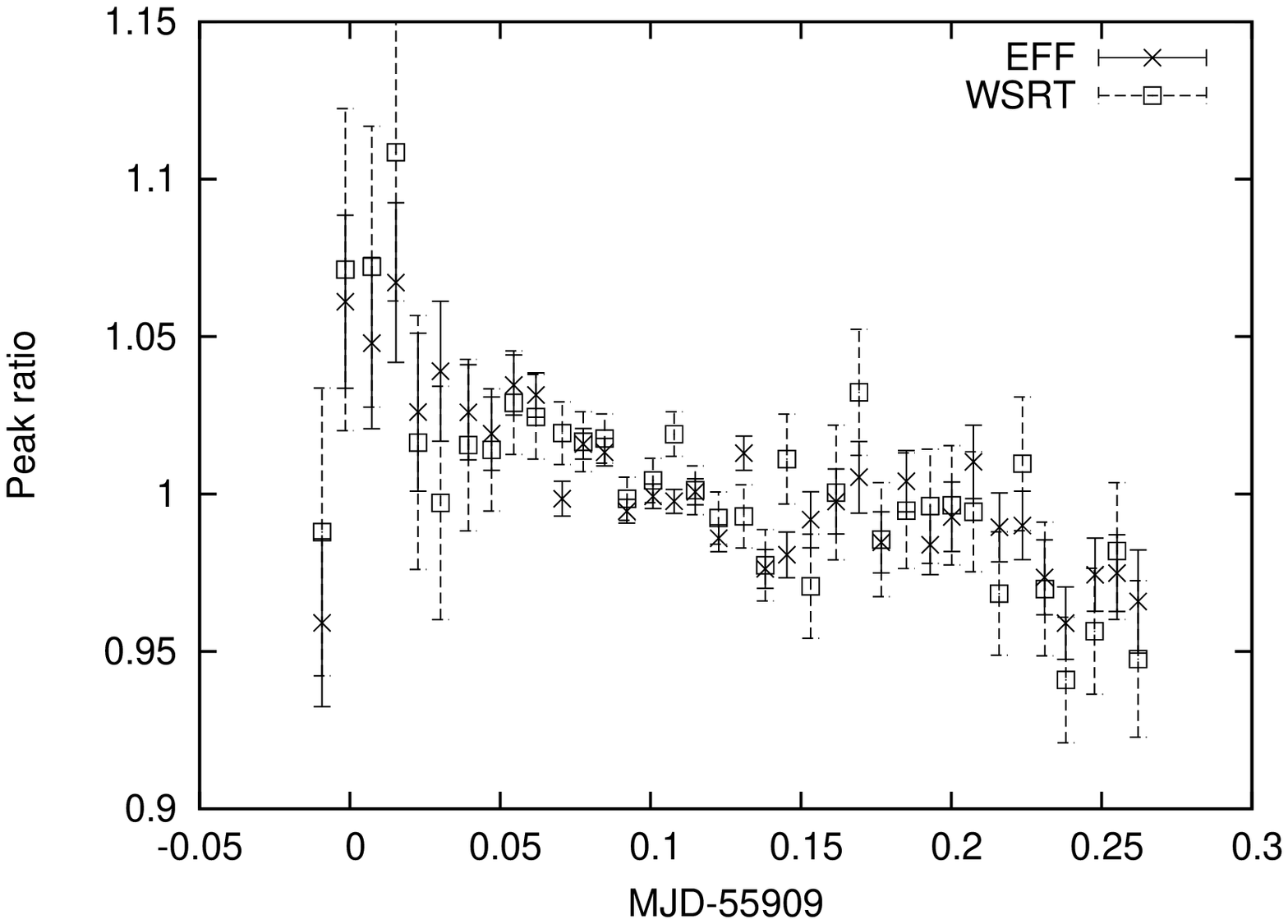}
\includegraphics[scale=0.48,angle=-90]{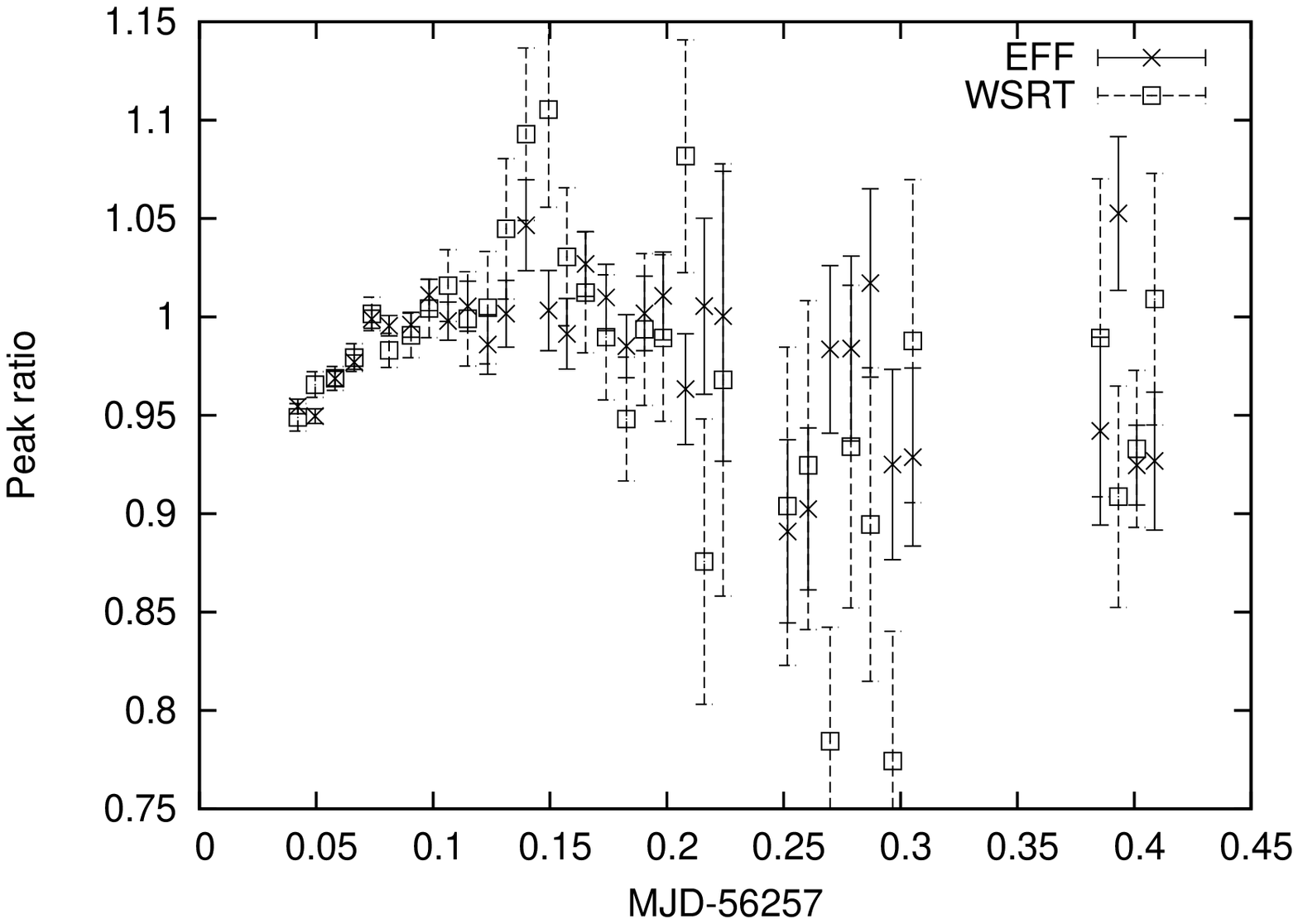}
\caption{Amplitude ratio measurements of the two components in the
PSR~J1022+1001 total intensity profiles from two epochs. The
measurements were made using 10-min integrations of a 40\,MHz band
centred at 1320\,MHz. Note that measurements of fractional error
larger than 10\% were discarded, which leads to the gap within the
results at MJD~56257. \label{fig:Pratio}}
\end{figure}

\begin{figure}
\centering
\includegraphics[scale=0.48,angle=-90]{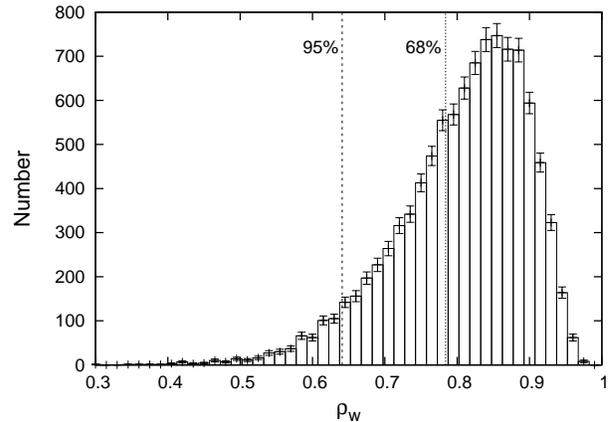}
\caption{Distribution of $10^4$ simulated values of $\rho_{\rm w}$
based on amplitude ratio measurements from data obtained at
MJD~56257. \label{fig:ccw_stat}}
\end{figure}

As mentioned in Section~\ref{sec:intro}, it has not been
conclusively shown whether the average profiles of PSR~J1022+1001
exhibit intrinsic shape variation. Here, following the idea of
previous analyses \citep[e.g.][]{kxc+99}, we calculate the amplitude
ratio of the leading and trailing components of total intensity
profiles for 10-min integrations, and plot them against time. For each component, the amplitudes were defined as the total
intensity within a fixed phase range (1.7\% of a period), after
subtracting the baseline. Their errors were calculated from the rms
of the off-pulse bins. We used a 40\,MHz sub-band ranging from 1300
to 1340\,MHz, where for most of our observing time, the flux density
is the highest within the entire overlapping band. The results from
data at MJD~55909 and 56257 are shown in Fig.~\ref{fig:Pratio},
where at both epochs the measurements show a clear evolution with
time and are consistent between the two sites. The variation trend
is similar to what has been indicated in Fig.~3.6 of \cite{pur10}.
We also calculated the weighted correlation coefficients (
$\rho_{\rm w}$s) between trends from the two sites, which is defined
as
\begin{equation}
\rho_{\rm w}=\frac{{\rm cov}_{\rm w}(\mathcal {E}_i,\mathcal
{W}_i,\sigma_{\mathcal {E},i},\sigma_{\mathcal {W},i})}{\sqrt{{\rm
cov}_{\rm w}(\mathcal {E}_i,\mathcal {E}_i,\sigma_{\mathcal
{E},i},\sigma_{\mathcal {E},i}){\rm cov}_{\rm w}(\mathcal
{W}_i,\mathcal {W}_i,\sigma_{\mathcal {W},i},\sigma_{\mathcal
{W},i})}}.
\end{equation}
Here $\mathcal {E}_i$, $\mathcal {W}_i$ are the ratio measurements
from Effelsberg and WSRT, respectively, $\sigma_{\mathcal {E},i}$,
$\sigma_{\mathcal {W},i}$ are their measurement errors, and ${\rm
cov}_{\rm w}$ is the weighted covariance defined by
\begin{equation}
{\rm cov}_{\rm w}(x_i,y_i,\sigma_{x,i},\sigma_{y,i})=
\frac{\displaystyle\sum_i(x_i-\overline{x})(y_i-\overline{y})/(\sigma^2_{x,i}\sigma^2_{y,i})}{\displaystyle\sum_i1/(\sigma^2_{x,i}\sigma^2_{y,i})},
\end{equation}
where the overline stands for weighted mean. The calculated
$\rho_{\rm w}$ is 0.60 at MJD~55909 and 0.91 at MJD~56257,
respectively, confirming that measurements from the two sites
exhibit the same evolution trend. Note that at the beginning of
MJD~56257 the pulsar was significantly brighter and the profile was
showing a steeper variation trend than at MJD~55909, which thus
corresponds to a value of $\rho_{\rm w}$ closer to unity. To
understand the estimation error of $\rho_{\rm w}$, we performed a
Monte Carlo simulation. In each iteration, we randomly altered the
measured component amplitudes based on Gaussian distributions with
the standard deviation equal to the measurement errors, and thus
calculated a new amplitude ratio and $\rho_{\rm w}$.
Fig.~\ref{fig:ccw_stat} shows the histogram of $\rho_{\rm w}$ with
10,000 iterations obtained from measurements at MJD~56257, where
$\rho_{\rm w}$ is larger than 0.78 and 0.64, with 68\% and 95\%
confidence level, respectively. The same investigation has also been
performed for the measurements at MJD~55909 and the corresponding
thresholds are 0.40 and 0.20, respectively. These mostly rule out
the possibility of a non-correlation between the two sites. A brief
summary of the results can be found in Table~\ref{tab:corr}.

\begin{table}
\centering \caption{Weighted correlation coefficient ($\rho_{\rm
w}$) between the amplitude ratio measurements from the WSRT and
Effelsberg data at two epochs, as well as their lower limits
with 68\% ($\mathcal {L}_{68}$) and 95\% ($\mathcal {L}_{95}$)
confidence level.} \label{tab:corr}
\begin{tabular}[c]{cccc}
\hline
MJD   &$\rho_{\rm w}$    &$\mathcal {L}_{68}$    &$\mathcal {L}_{95}$ \\
\hline
55909 &0.60              &0.40                   &0.20   \\
56257 &0.91              &0.78                   &0.64   \\
\hline
\end{tabular}
\end{table}

\begin{figure}
\centering
\includegraphics[scale=0.35,angle=-90]{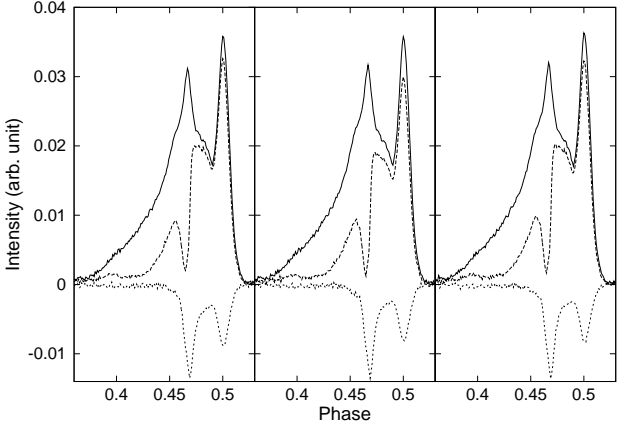}
\caption{Comparison of Effelsberg polarisation profiles obtained
from data at MJD~56257, based on calibrator+single-axis model
(left), template+reception model (middle), and template+single-axis
model (right). The solid, dashed and dotted lines stand for the
Stokes parameter $I$, $L$ and $V$, respectively. \label{fig:cals}}
\end{figure}

To investigate the effect of polarisation calibration on our
results, we also applied the template-matching calibration technique
to the Effelsberg data at MJD~55909 and 56257 \citep[][Lee et al.~in
prep.]{van06}, based on the template polarisation profile from
European Pulsar Network Pulse Profile Database \citep{stc99} showing
consistent polarimetry with previous published results. For this
calibration scheme, we used both the single-axis and the reception
description of the receiver properties, the latter of which models
both the differential gain and cross-coupling of the two feeds
\citep{ovhb04,van04a}. The calibrated data were then processed by
following the same procedure as described above. Both receiver
models led to a fit with reduced $\chi^2$ close to unity. Still, the
effect of feed cross-coupling was estimated to influence the total
intensity profile by less than 1\%, and the true value could
not be measured due to limited sensitivity. As a demonstration, in
Fig.~\ref{fig:cals} we show the obtained polarisation profiles from
MJD~56257 data with the calibration result based on calibrators and
the single-axis model shown in Fig.~\ref{fig:prof-cal}. It can
be seen that the three profiles from different calibration schemes
or models exhibit consistent polarisation properties, especially the
two achieved with the single-axis model. Using the reception model
results in a slightly lower linear polarisation percentage (middle
panel), mostly due to absorption of power into the feed leakage
terms. In addition, the calculated $\rho_{\rm w}$s between the new
amplitude ratio measurement trend and the WSRT result, is 0.89 when
using the reception model and 0.91 when single-axis calibration was
applied. These are almost the same as the value obtained above. The
same analysis at MJD~55909 data led to $\rho_{\rm w}=0.55$ when
using the reception model and $\rho_{\rm w}=0.60$ when applying
single-axis calibration, also consistent with the result based on
calibrators. Further investigation based on independent Effelsberg
data for the Large European Array for Pulsars project \citep{ks10},
shows that the circular receiver does not suffer from strong
cross-coupling between the two feeds (Lee et al.~in prep.), which
validates the use of the single-axis model in this case. Therefore,
it is highly unlikely that our detected profile variation is
due in large part to improper polarisation calibration.

\begin{figure}
\centering
\includegraphics[scale=0.48,angle=-90]{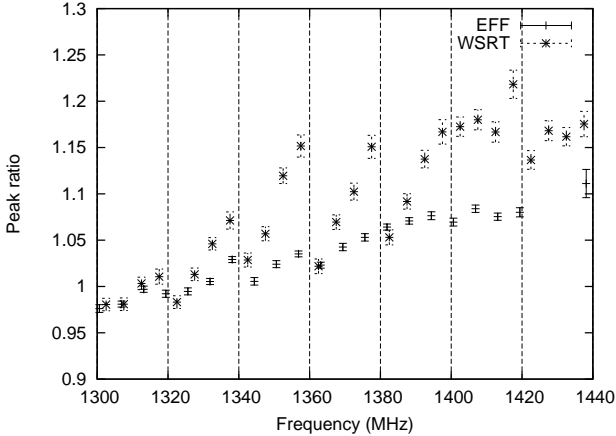}
\caption{Measured amplitude ratios as a function of frequency, from
the profile models obtained at MJD~55909. The measurements from the
Effelsberg and WSRT data are based on profiles of bandwidth 6.25 and
5\,MHz, respectively. The edges of the 20\,MHz sub-bands
(starting from 1300\,MHz) in the WSRT data have been marked by
dashed lines. Note that for the WSRT data there is an additional
increase of the measurement values within a sub-band, associated
with the low-bit digitisation artefact shown in
Fig.~\ref{fig:prof-cal}. More details of this effect are shown in
Appendix~\ref{sec:app}. \label{fig:PR-chan}}
\end{figure}

\begin{figure*}
\centering
\includegraphics[scale=0.48,angle=-90]{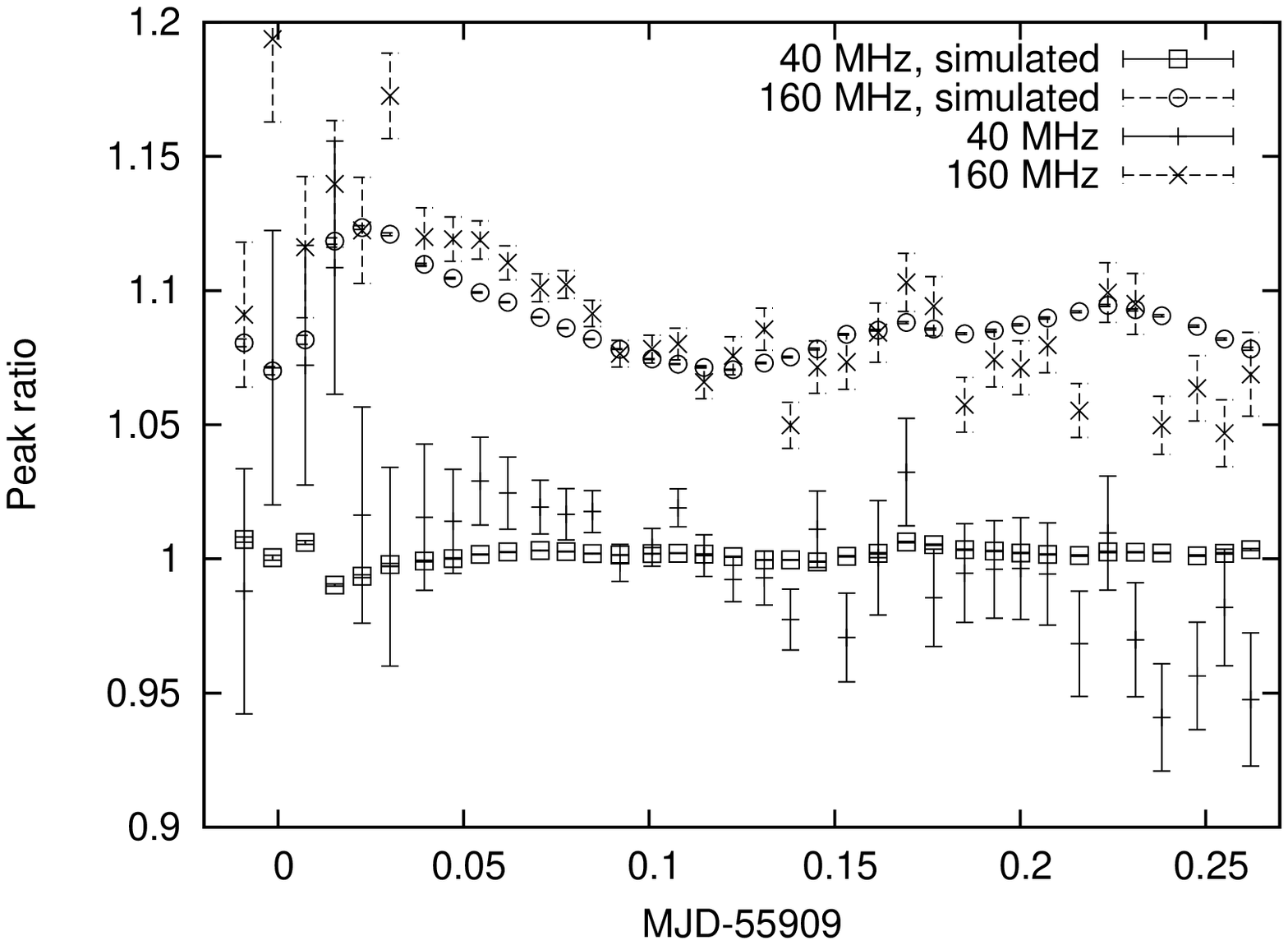}
\includegraphics[scale=0.48,angle=-90]{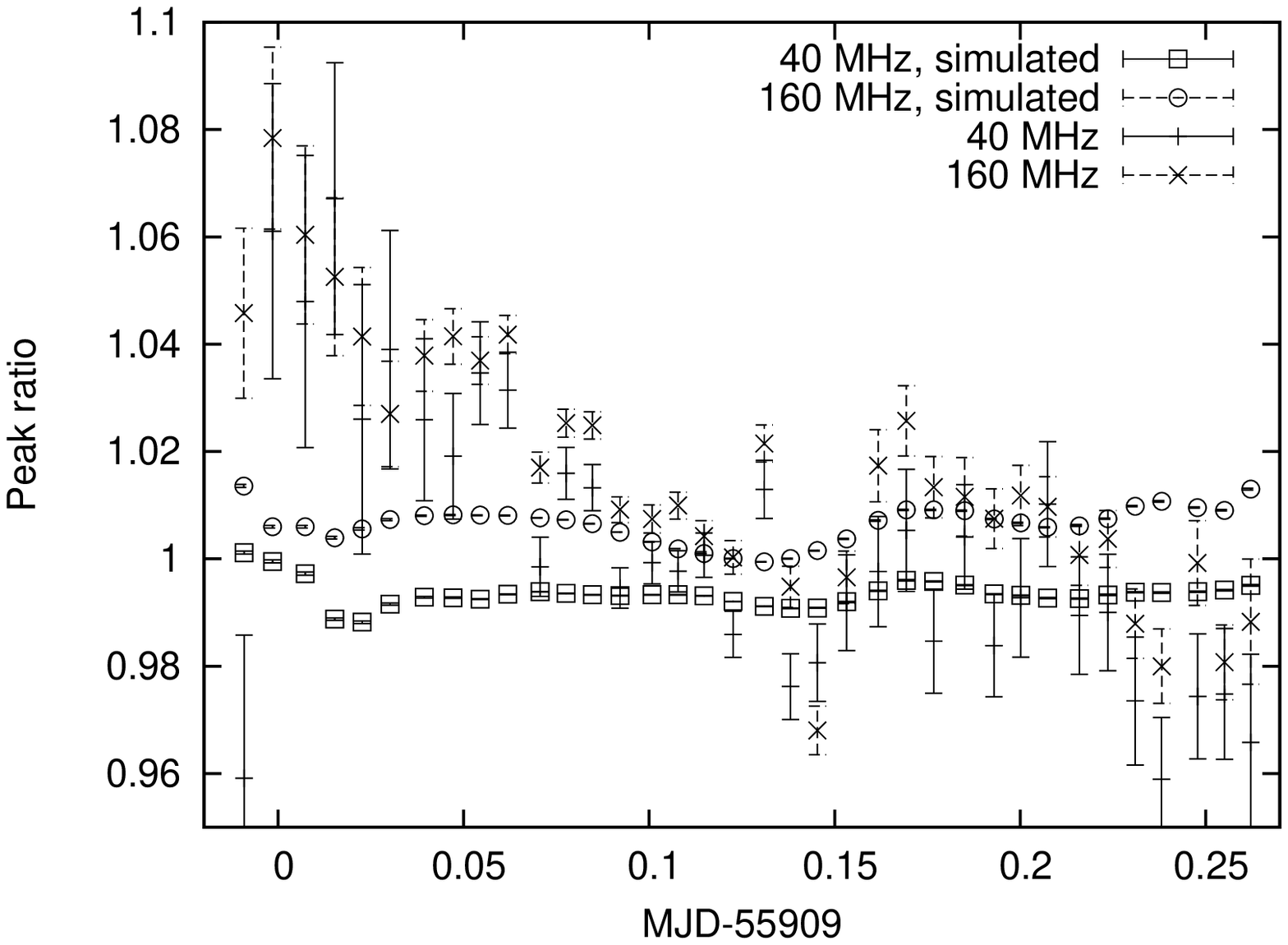}
\includegraphics[scale=0.48,angle=-90]{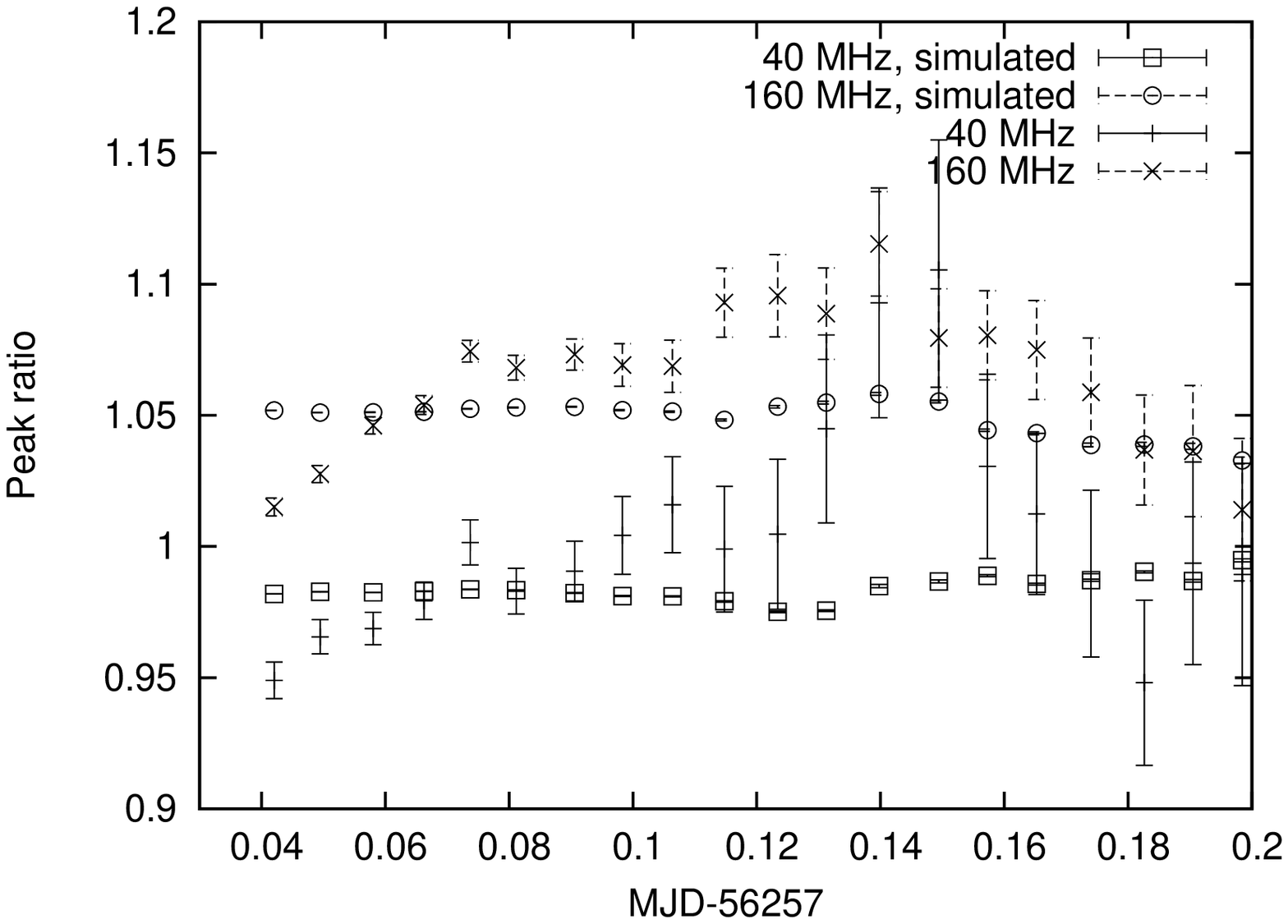}
\includegraphics[scale=0.48,angle=-90]{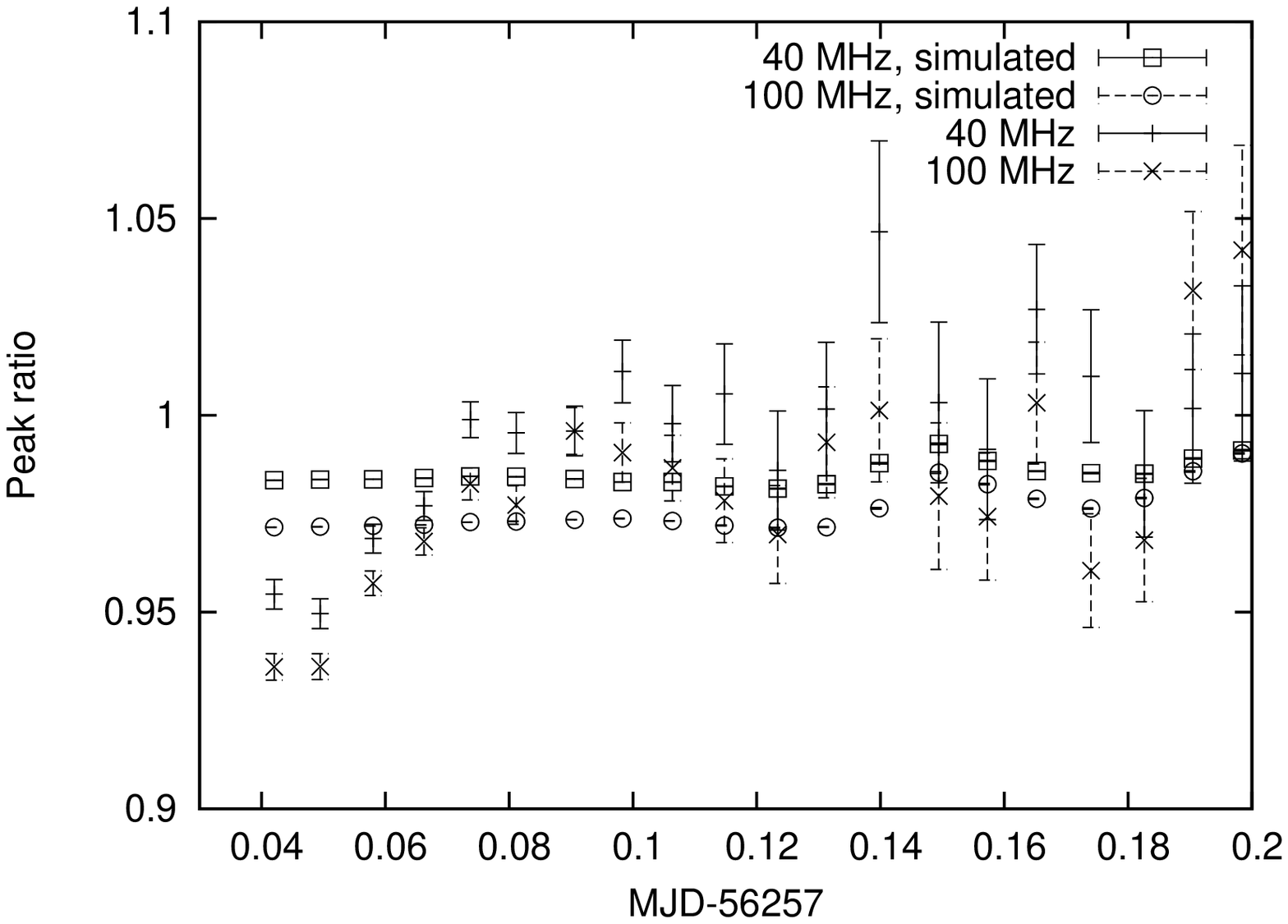}
\caption{Simulated peak amplitude ratios based on flux densities
measured from 10-min integrations, compared with the actual
ratio measurements. The simulations were carried out on both the
WSRT (left) and Effelsberg (right) data at MJD~55909 (top) and 56257
(bottom), using both the 40\,MHz sub-band and the entire available
bandwidth. The uncertainties of the simulated ratios are typically
of order $\lesssim10^{-3}$, and thus not visible from plots. Note
that for MJD~56257, the analysis was carried out only for the first
half of the data, as in the second half the flux densities within
individual frequency channels were not measurable on many occasions
due to low S/Ns. The effective observing bandwidth of Effelsberg at
MJD~55909 is 160\,MHz due to a receiver cutoff below
1280\,MHz. \label{fig:simuPratio}}
\end{figure*}

Note that the profile shape has a clear frequency dependence
\citep[e.g.][]{rk03}. If the flux density distribution within the
observing band changes significantly in time due to ISM diffractive
scintillation, the averaged profiles (after frequency summation) may
therefore also exhibit time-variant behaviour
\citep[e.g.][]{lvk+11}. To understand its contribution to our
results, here we simulate the varying trends by this effect and
compare them with the real measurements. At each epoch, we first
used an integrated profile for the entire duration of the
observation to establish a profile model with frequency dependent
shape. As an example, Fig.~\ref{fig:PR-chan} shows the measured
amplitude ratios as a function of frequency\footnote{The defined
phase ranges to calculate component amplitudes are identical for all
central frequencies as the data were already de-dispersed as
mentioned in Section~\ref{sec:obs}.} from the profile models for
the MJD~55909 data. We then applied the measured
frequency-dependent flux densities from each 10-min integration to
the model to create a simulated profile, and computed the resulting
amplitude ratio after frequency summation. For this purpose, we
kept the frequency resolution of 6.25\,MHz for the Effelsberg data
and 10\,MHz for the WSRT data, so as to have enough signal within
each frequency channel. The channel width is significantly smaller
than the estimated scintillation bandwidth of this pulsar
\citep{cl02,yhc+07}. The uncertainties in the amplitude ratios
of the simulated profiles were calculated from the measurement
errors of the flux densities estimated from each frequency channel
of each 10-min integration. The flux densities as well as their
errors were measured by using \textsc{pdv} (psrchive's archive data
displayer), where the flux density is defined as the total intensity
within a pulse period after subtracting the baseline, and their
errors were calculated from the rms of off-pulse phase bins defined
by 3-$\sigma$ threshold. By following this procedure, the observed
variation trend would be reproduced if the profile instability is
mostly dominated by the scintillation effect. In
Fig.~\ref{fig:simuPratio} we present the results at MJD~55909 and
56257 from both sites, based on both the selected 40\,MHz sub-band
and the entire bandwidth. Clearly, in most cases the simulation
resulted in a flat trend and did not reproduce the actual
measurements\footnote{Note that an exceptional case may be drawn
from the simulation on full bandwidth of the WSRT data from
MJD~55909. This indicates that profile evolution based on large
bandwidth could be significantly contributed by scintillation
effect. The reason why the simulation based on Effelsberg data from
the same epoch did not reproduce the variation trend, is that our
Effelsberg profile shows significantly less frequency
dependency and is thus less affected by the scintillation
effect.}. The reduced $\chi^{2}$s of the measurement series with
respects to their corresponding simulated trend\footnote{Defined as
$\displaystyle\frac{1}{N_{\rm
r}-1}\sum_{i}\frac{(r_i-s_i)^2}{\sigma^2_{r,i}+\sigma^2_{s,i}}$,
where $N_{\rm r}$ is the number of amplitude ratios, $r_i$, $s_i$
are amplitude ratios from real measurement and simulation,
respectively, and $\sigma_{r,i}$, $\sigma_{s,i}$ are their errors.}
fell in between 2.0 and 22, all significantly above unity. We
therefore conclude that our detected profile instability shown in
Fig.~\ref{fig:Pratio} is not significantly affected by the
scintillation effect.

We do not detect profile variation at the other two epochs,
MJD~55974 and 56158, mostly because the source was comparatively dim
due to scintillation. At these two epochs, with the WSRT
160\,MHz observations the S/N$_{\rm 5 min}$ values are all below 30,
while the maxima at MJD~55909 and 56257 reach 75 and 110,
respectively. This may explain the contradictory conclusions from
previous studies on profile stability on short timescales. In
addition, it can be seen from Fig.~\ref{fig:Pratio} that the
amplitude ratio variation may not be significant enough for
detection within 0.1\,day, which is the longest observation used
in previous work. Therefore, detection of profile instability
is achievable only when the pulsar is bright and the variation is
large.

We note that our analysis draws seemingly different
conclusions on the profile stability of PSR~J1022+1001 compared to
\cite{lkl+11}, due to a few reasons. Firstly, our new observations
detected the pulsar when it was significantly brighter than in
the previous ones. For instance, at the beginning of the MJD~56257
observation an integration time of 1800\,s led to a profile with signal-to-noise ratio over 800, while integrating the entire
observation (approximately 1,900\,s) in \cite{lkl+11} only resulted
in a value of about 480. Secondly, the work in \cite{lkl+11} was
based on 1-min integrations and thus focused more on searching for
random variations in profile shapes on short timescales. Finally,
the duration of the previous observation does not provide the
sensitivity to profile variations on a timescale of several tens of
minutes detected in this paper.

\begin{figure}
\centering
\includegraphics[scale=0.35,angle=-90]{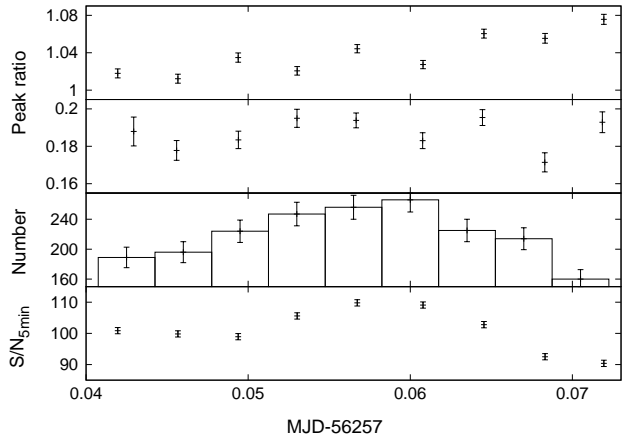}
\caption{Top to bottom: amplitude ratio measurements from 5-min
integrations and from sub-pulse averages, numbers of sub-pulses in
5-min, S/Ns of 5-min integrations. Here the sub-pulses were selected
with $({\rm S/N}_{\rm s})/(\langle{\rm S/N}_{\rm s}\rangle_{\rm
5min})>9$ and MJD between 56257.042 and 56257.074. The 5-min
integrations were formed from WSRT data. \label{fig:PR-sgl}}
\end{figure}

In principle, the profile variation can be fully attributed to the
detected bright sub-pulses, if either their shape or occurrence rate
varies by a sufficiently large amount in time. Such a possibility
has been investigated in Fig.~\ref{fig:PR-sgl}. To balance between
potential shape bias introduced by a variable detection threshold
and time span allowing us to see profile variation, here we selected
sub-pulses detected with $({\rm S/N}_{\rm s})/(\langle{\rm
S/N}_{\rm s}\rangle_{\rm 5min})>9$ and MJD between 56257.042 and
56257.074. It can be seen that while the amplitude ratios from 5-min
integrations is gradually increased in time ($\approx6$\% from
MJD~56257.04 to 56257.07), no similar variation trend is witnessed
from the averages of the sub-pulses. This indicates that the profile
variation is not caused by the selected bright sub-pulses. In fact,
the number of detected sub-pulses is approximately $200-300$ per
5-min window ($1-1.5\%$ of all periods). Therefore, the amplitude
ratios of the sub-pulse averages would need to change by a factor of
4 or more if variability in the $1-1.5\%$ of detected
sub-pulses were the cause of the profile variation in 5-min
integrations. Note that the number of detected sub-pulses
remain consistent in time, though it is proportional to the S/N
of 5-min integrations, which is expected due to variability in
detection sensitivity. This suggests that the profile variation is
not induced by the occurrence rate variation of bright sub-pulses.
In fact, assuming profile variation mainly occurs at the trailing
component, the increase of amplitude ratio from 5-min integrations
corresponds to 6\% decrease of the flux density at the trailing
component. Given the average amplitude ratio of 0.18 observed
in the detected sub-pulses and an amplitude ratio of 1 for ordinary
integrations, the selected sub-pulses contribute 6-9\% of the flux
density at the trailing component. Therefore, if the change of
occurrence rate of the selected sub-pulses were fully responsible
for the amplitude ratio variation, the number of detections around
MJD~56257.07 would have dropped by 70-100\%, which is clearly not
the case. Therefore, it is unlikely that our selected sub-pulses
cause the detected profile variation, which suggests that sub-pulse
variability and integrated profile variation are two different
effects. Nevertheless, the analysis is still restricted to roughly
the brightest 1\% of all single pulses, and will be significantly
improved by increasing system sensitivity.

\subsection{Timing of sub-pulses} \label{ssec:timing}
To investigate the impact of sub-pulses on pulsar timing, we used
the sub-pulses selected in Fig.~\ref{fig:widvf} (3.4\% of all
periods, from the trailing component, in total approximately 4,500
pulses) and compared their timing with ordinary integrations based
on all rotations. The TOAs are measured by following the classic
template-matching method described in \cite{tay92}. The template
profile for timing based on sub-pulses is formed by summing all
detected sub-pulses and performing a Gaussian component smoothing as
in e.g. \cite{kra94}. The timing residuals were calculated with the
\textsc{TEMPO2} software package \citep{hem06}. Here we used the
timing solution obtained from the European Pulsar Timing Array
collaboration (Desvignes et al. in prep.), without fitting for any
parameters.

\begin{figure}
\centering
\includegraphics[scale=0.5,angle=-90]{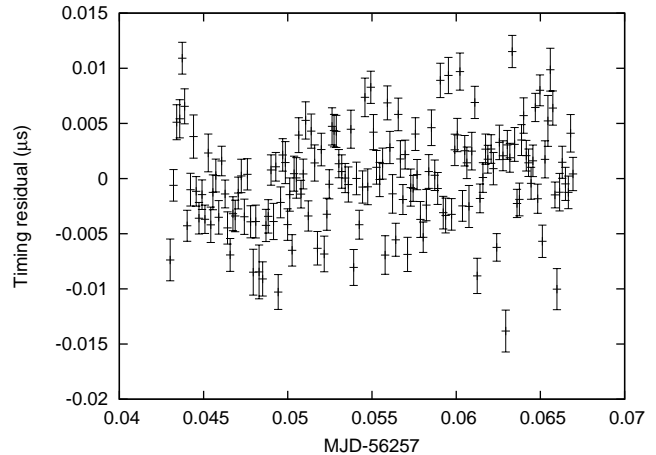}
\caption{Timing residuals based on 151 averages of 30 sub-pulses.
The rms is 4.3\,$\mu$s while the white noise rms indicated from the
TOA errors is 1.6\,$\mu$s. \label{fig:res}}
\end{figure}

\begin{figure}
\centering
\includegraphics[scale=0.5,angle=-90]{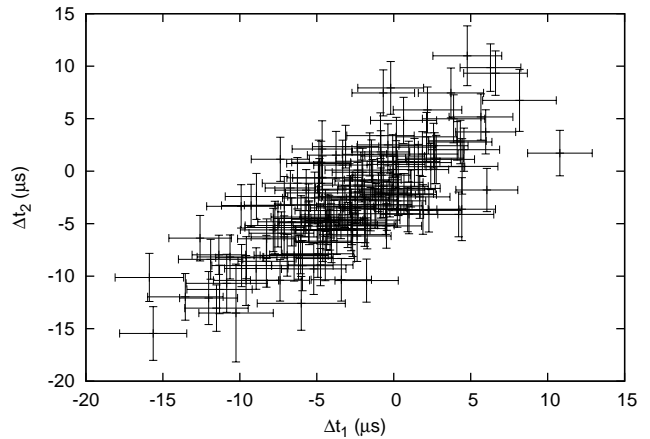}
\caption{Correlation plot of TOA pairs calculated based on sub-pulse
averages from two individual 80\,MHz sub-bands. The data used here
are the same as in Fig.~\ref{fig:res}. \label{fig:bd12}}
\end{figure}

In Fig.~\ref{fig:res} we present the timing residuals calculated
with averages of 30 pulses (not contiguous in time), equivalent to
an integration time of 0.5\,s. The timing solution has an rms of
4.3\,$\mu$s which is greater than the average of the errors owing to
white noise (1.6\,$\mu$s). This suggests the existence of phase
jitter. To better understand the timing noise, a standard
Kolmogorov-Smirnov (K-S) test has been performed on the TOAs as done
in \cite{lkl+11}. The measured p-value ($\simeq0.98$) is close to 1,
indicating no significant deviation of the residuals from a Gaussian
distribution \citep[e.g.][]{pftv86}. In Fig.~\ref{fig:bd12} we
divide the entire band into two sub-bands, and present a correlation
plot of TOAs from them, as in \cite{ovh+11} and \cite{sc12}. The TOA
pairs are shown to be highly correlated and the rms of each
individual group is also estimated to be roughly equal
(4.6\,$\mu$s), meaning that the noise is correlated and dominated by
phase jitter.

In contrast, we did not detect any significant impact of pulse phase
jitter on timing with ordinary integrations. Data obtained between
MJD~56257.044 and 56257.080, when the observing S/N was maximal,
have been examined based on 1-min averages. The resulting timing rms
is 1.3\,$\mu$s, close to the estimated white noise level
(1.1\,$\mu$s) from the TOA errors. Subtraction of the white noise
contribution from the timing residuals leads to an upper limit of
$\sim700$\,ns for jitter noise based on a 1-min integration time.

\begin{figure}
\centering
\includegraphics[scale=0.5,angle=-90]{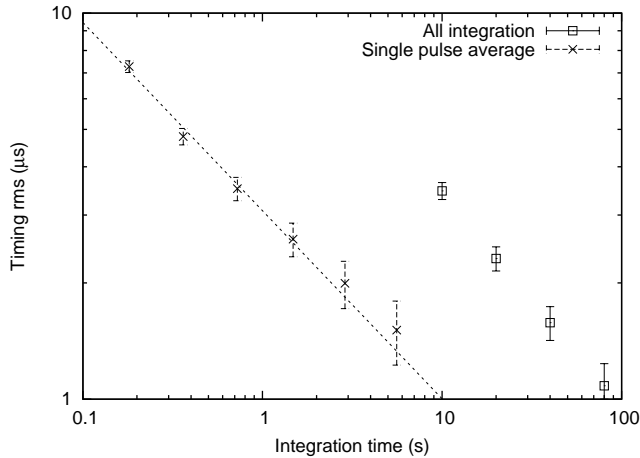}
\caption{Timing rms of averages based on different numbers of bright
sub-pulses, compared with rms from integrations of all periods. The
dashed line shows a power-law modelling with a fitted index of
$-0.49\pm0.02$. Note that an integration of 10\,s corresponds to
roughly 600 pulses. \label{fig:rms}}
\end{figure}

In Fig.~\ref{fig:rms} we present the timing rms values when altering
the number of integrated sub-pulses and compare them with timing rms
yielded by ordinary integrations. It is shown that the jitter noise
decreases following the $\sigma_{\rm J}\propto t_{\rm int}$ law as
expected \citep{cs10}, where $t_{\rm int}$ is the integration length
of each average. Averaging 30 sub-pulses leads to 151 TOAs of rms
4.3\,$\mu$s. Considering that during the same period of time
ordinary 1-min integrations lead to 34 TOAs of rms 1.3\,$\mu$s,
timing the averages of sub-pulses is almost as effective as
timing the ordinary integrations
($4.3/\sqrt{151/34}\simeq2.0$\,$\mu$s, close to 1.3\,$\mu$s). Note
that only 3.4\% of the single rotations were picked up, an improved
system sensitivity would enable significantly more sub-pulse
detections. Including those sub-pulses in timing may significantly
decrease the jitter noise and make the timing based on sub-pulses
more efficient (achieving more measurements of the same rms in the
same observing time, or equal number of measurements with lower rms)
than using all-period integrations. However, this extrapolation is
based on the assumption that the sub-pulses to be included exhibit
the same phase jittering as the currently selected ones, which
requires further investigations with improved sensitivity to
validate.

\section{Conclusion and discussion} \label{sec:conclu}
In this paper, we have performed a detailed investigation into the
profile variation of PSR~J1022+1001. Within our 35-hr observations,
approximately 14,400 sub-pulses were detected above a 5-$\sigma$
threshold from the WSRT baseband data, and most of them are
coincident with the trailing component of the integrated profile.
The flux densities and polarisation properties suggest that they are
the bright end of the pulse energy distribution and not a separate
population. No giant pulse has been detected. Pulses from the
leading and trailing components can happen within the same rotation
and their occurrences are shown to be correlated. For pulses from
the trailing components, a preferred pulse width of approximately
0.25\,ms has been found. Using simultaneous observations at the
Effelsberg telescope and the WSRT, we have tracked the variation of
integrated profiles on a timescale of several tens of minutes.
Profile instability due to improper polarisation calibration and
diffractive scintillation was excluded. We have not yet discovered
an association between integrated profile instability and sub-pulse
occurrence, possibly owing to the present sensitivity limit. In
addition, we have demonstrated the dominance of phase jitter in
timing residuals with the sub-pulse averages from the trailing
component, and obtained an upper limit of $\sim700$\,ns for jitter
noise based on continuous 1-min integrations. Further investigation
with better sensitivity is still required to find out if timing the
averages of sub-pulses is more efficient than using continuous
integrations.

The spin-down power ($\dot{E}$) of PSR~J1022+1001 is estimated to be
$3.8\times10^{32}$\,$\rm ergs~s^{-1}$ \citep[obtained from ATNF
Pulsar Catalog, for details see][]{mhth05}, significantly less than
those of the current few MSPs with giant pulse discoveries
\citep{cst+96,rj01,jkl+04,kbmo05}. Thus, our non-detection of giant
pulses from PSR~J1022+1001 tends to coincide with the correlation
between giant pulse emissivity and spin-down luminosity in MSPs
\citep[e.g.][]{kbmo05}. Still, the detected sub-pulses satisfy the
criterion of the so-called ``giant micropulses'', which requires the
peak flux densitiy much greater than 10 times that of the average
profile, but the integrated flux density not exceeding 10 times the
average \citep{cai04}. Our detection, together with the discovery in
PSR~J0437$-$4715 \citep{jak+98}, suggests that emissivity of giant
micropulses in MSPs is not necessarily related to spin-down
luminosity. This suggests the possibility that giant
micropulses may have a different origin from giant pulses.

The two components of PSR~J1022+1001 are separated by only 5\% of
the period, but have dramatically different emission
behaviours. It could be that the two components actually
originate from well-separated regions in height, due to the strong
winding of the magnetosphere \citep[e.g.][]{pet11}. A fit of the
aberration \& retardation field-line model to the integrated profile
which estimates the emission latitude, may shed more light on this
issue \citep{drh04}.

Admittedly, single pulse studies of MSPs are still mostly limited by
detection sensitivity. With the next generation of radio telescopes,
such as the Five-hundred-metre Aperture Spherical Radio Telescope
and the Square Kilometre Array, the system sensitivity for pulsar
observations will be increased by 1-2 orders of magnitude from the
current level \citep[e.g.][]{lvk+11}. This will enable more
comprehensive investigations into MSP single pulses, by increasing
the number of detections, improving the detection qualities, and
enabling more case studies. Such efforts would be greatly helpful in
understanding the behaviour of single pulse emission as well as its
impact on pulsar timing stability, and thus developing potential
approaches to improve timing precision.

\section*{Acknowledgements}
We thank J.~P.~W.~Verbiest for sharing the ephemeris for our timing
analysis, and are grateful to A.~Jessner and L.~Guillemot for
valuable discussions. We would also like to thank the anonymous
referee who provided constructive suggestions to improve the paper.
K.~Liu is supported by the ERC Advanced Grant ``LEAP", Grant
Agreement Number 227947 (PI M.~Kramer). This work was carried out
based on observations with the 100-m telescope of the
Max-Planck-Institut f\"{u}r Radioastronomie at Effelsberg. The
Westerbork Synthesis Radio Telescope is operated by the Netherlands
Foundation for Radio Astronomy, ASTRON, with support from The
Netherlands Foundation for Scientific Research (NWO).

\bibliographystyle{mnras}
\bibliography{journals_apj,psrrefs,modrefs,crossrefs}

\appendix
\section{Example of corrupted data due to faulty power level setting} \label{sec:app}
The data from the failed observing session at MJD~55974 at the
Effelsberg telescope provide insights into the effect of 2-bit
sampling on profile evolution. In this observing run, the signal
level attenuator was inadvertently set too high, resulting in
voltages that were too small as input to the 8-bit
analog-to-digital converter (ADC) and reducing it to a low-bit
(2-bit or even smaller) system. This data allow us to assess the
effect of low-bit sampling on PSR~J1022+1001 profile shape
variation, thus highlighting an extreme case of signal clipping.

\begin{figure}
\centering
\includegraphics[scale=0.4,angle=-90]{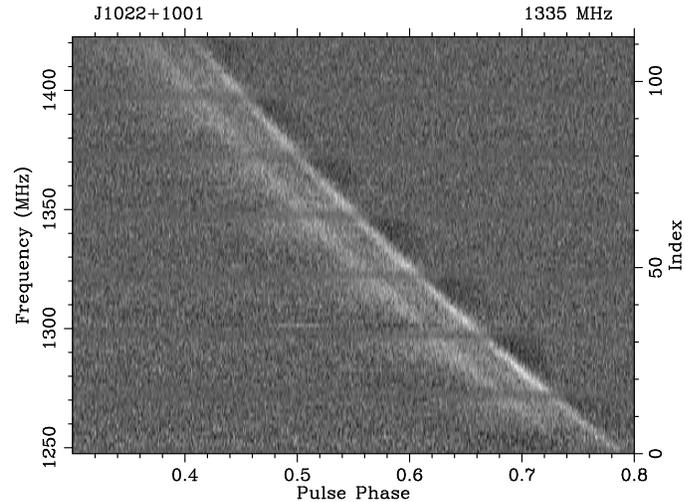}
\caption{Gray-scale image of the profile against frequency obtained
from the Effelsberg at MJD~55974 with 4-hr integration
time.\label{fig:GTp}}
\end{figure}

\begin{figure}
\centering
\includegraphics[scale=0.37,angle=-90]{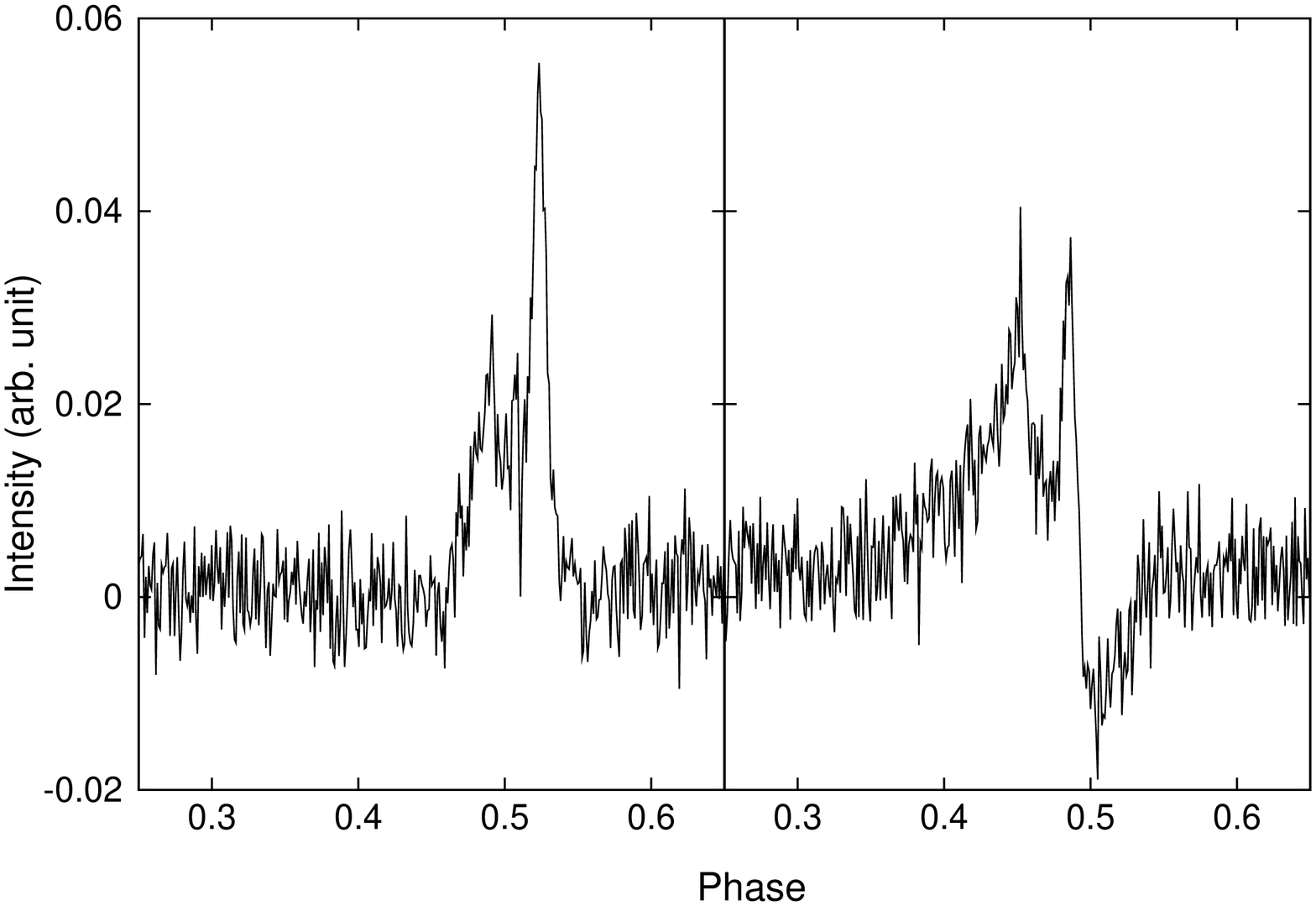}
\caption{Pulse profiles formed with a bandwidth of 3\,MHz at the
lower (left) and upper (right) side of the 1285\,MHz sub-band,
respectively, from the data shown in Fig.~\ref{fig:GTp}. The central
frequencies are 1277 and 1293\,MHz, respectively. \label{fig:p12}}
\end{figure}

Fig.~\ref{fig:GTp} shows a gray-scale image of the profile as a
function of frequency before dedispersion. Within each 25\,MHz
sub-band, there is negative power distributed in a similar fashion
as shown by Fig.~4 of \cite{ja98}, which is a consequence of low-bit
sampling. As the pulsar signal is dedispersed, the negative dip
which would suppress the power of the signal, leads the profile at
lower frequencies and trails at higher ones. This will result in
additional peak amplitude ratio variation across the sub-band which
can be witnessed in Fig.~\ref{fig:p12}. Here, the profile from the
lower side of a sub-band exhibits a significantly suppressed
first component, while that from the upper side is affected
mostly at the phase of the trailing component.

\begin{figure}
\centering
\includegraphics[scale=0.5,angle=-90]{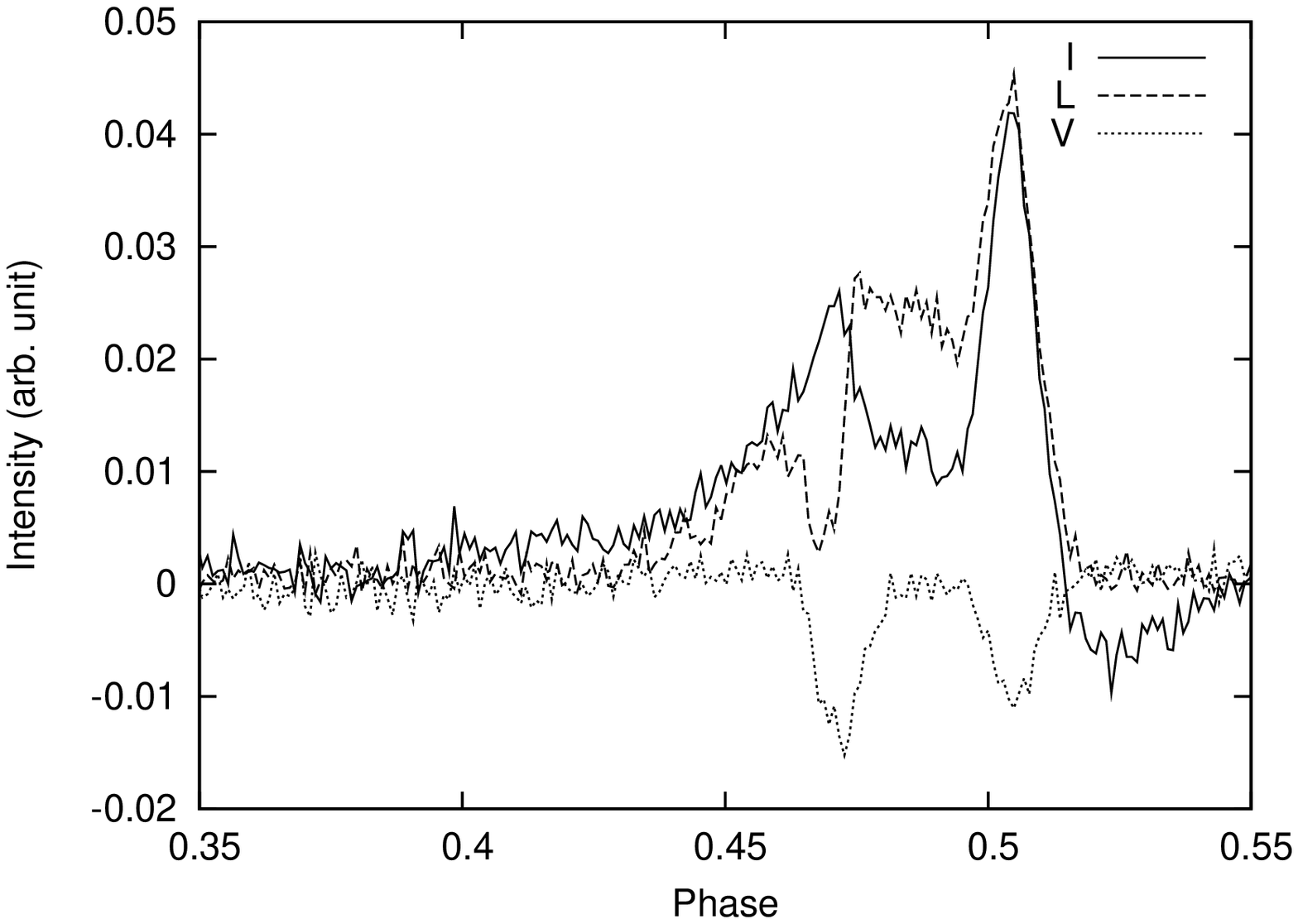}
\caption{Polarisation profile of the data in Fig.~\ref{fig:GTp},
averaging over all frequencies. \label{fig:pol}}
\end{figure}

Fig.~\ref{fig:pol} shows the polarisation profile of data in
Fig.~\ref{fig:GTp}. It can be seen that the linear and circular
components in general retain their original shape, while the total
intensity profile is greatly suppressed and lower than the
linear component at most of the on-pulse phases including the valley
region between the two main peaks. This explains why the WSRT
profile shown in Fig.~\ref{fig:prof-cal} is slightly over-polarised.

\end{document}